\documentclass[12pt]{article}
\usepackage[utf8]{inputenc} 
\usepackage[T1]{fontenc}    
\usepackage{hyperref}       
\usepackage{url}            
\usepackage{booktabs}       
\usepackage{amsfonts}       
\usepackage{amsmath}
\usepackage{graphicx}
\usepackage{enumitem}
\usepackage{indentfirst}
\usepackage{multirow}
\usepackage{libertine}
\usepackage{gensymb}
\usepackage{algorithm} 
\usepackage{algpseudocode} 
\usepackage[english]{babel}
\usepackage[top=2.5cm, bottom=2.5cm, left=2.5cm, right=2.5cm]{geometry}
\usepackage{natbib}
\usepackage{caption}
\usepackage{subcaption}

\newtheorem{definition}{Definition}

\DeclareMathOperator*{\argmax}{arg\,max}

\usepackage{color}

\def\F{\boldsymbol{F}}
\def\sbf{\boldsymbol{s}}
\def\Cbf{\boldsymbol{C}}
\def\u{\boldsymbol{u}}
\def\w{\boldsymbol{w}}
\def\x{\boldsymbol{x}}
\def\X{\boldsymbol{X}}
\def\ctil{\tilde{c}}
\def\ytil{\tilde{y}}

\def\bbf{\boldsymbol{\beta}}
\def\gbf{\boldsymbol{\gamma}}
\def\thbf{\boldsymbol{\theta}}
\def\thatbf{\widehat{\thbf}}
\def\Sbf{\boldsymbol{\Sigma}}

\def\Fhat{{\widehat F}}

\title{Predicting Times to Event Based on Vine Copula Models}
\author{
	Shenyi Pan\thanks{Department of Statistics, University of British Columbia, Vancouver, BC Canada V6T 1Z4. Email: \href{mailto:shenyi.pan@stat.ubc.ca}{shenyi.pan@stat.ubc.ca}.}
	\and
	Harry Joe\thanks{Department of Statistics, University of British Columbia, Vancouver, BC Canada V6T 1Z4. Email: \href{mailto:harry.joe@ubc.ca}{Harry.Joe@ubc.ca}.}
}
\date{}

\begin{document}

\maketitle

\begin{abstract}
In statistics, time-to-event analysis methods traditionally focus on the estimation of hazards. In recent years, machine learning methods have been proposed to directly predict the event times. We propose a method based on vine copula models to make point and interval predictions for a right-censored response variable given mixed discrete-continuous explanatory variables. Extensive experiments on simulated and real datasets show that our proposed vine copula approach provides a decent approximation to other time-to-event analysis models including Cox proportional hazards and Accelerate Failure Time models. When the Cox proportional hazards or Accelerate Failure Time assumptions do not hold, predictions based on vine copulas can significantly outperform other models, depending on the shape of the conditional quantile functions. This shows the flexibility of our proposed vine copula approach for general time-to-event datasets.

\end{abstract}

Keywords: Survival analysis, time-to-event analysis, conditional quantiles, vine copula, copula regression, prediction interval.

\section{Introduction} \label{sec: intro}


In time-to-event and survival studies, the goal is to model the response variable $Y$ (i.e., the time-to-event) with the explanatory variables $\X = (X_1, \dots, X_p)$ to get point and interval predictions. The response $Y$ can be right-censored and the explanatory variables are random in an observational study.
A binary treatment variable can be accommodated if there is randomization to two treatments. The explanatory variables can be unbounded so that locally-based linear
approximations can be poor. For an observational study, a natural approach is fitting a joint distribution for $(X_1,\dots,X_p, Y)$ assuming a random sample $(x_{i1}, \dots, x_{ip}, y_i)$ for $i = 1, \dots, n$, and then obtain the predictive distribution for $Y$ given $\X=\x$.

Linear regression methods for time-to-event response data include the
semi-parametric Cox proportional hazards model, and parametric methods such as
Weibull regression and other accelerated failure time (AFT) models.
The focus often is on finding significant explanatory variables and
not on assessing quality of predictions.

In the past several decades, machine learning methods have also been proposed to predict a censored time-to-event response. Survival trees (\cite{gordon1985tree}) are a type of regression trees tailored to predict a censored response using a different splitting criterion. Random survival forest (\cite{ishwaran2008random}) extends survival trees by using a forest of these trees. \cite{fard2016bayesian} combine Bayesian methods including Naive Bayes, Tree-Augmented Naive Bayes, and Bayesian Network with the AFT model to predict event times using training data obtained at an early stage of the study. \cite{biganzoli1998feed} use the partial logistic artificial neural network to analyze the relationship between the explanatory variables and the event times. \cite{khan2008support} propose a support vector regression based method and adapt it for a censored response by using an updated asymmetric loss function. \cite{hothorn2006survival} extend gradient boosting machines to minimize a weighted risk function for time-to-event data. 
A review of methods is summarized based on \cite{wang2019machine}.
However, these machine learning methods only focus on point prediction and do not indicate how to get interval predictions. It can be difficult to obtain prediction intervals based on these methods.

In recent years, the vine pair-copula construction has proven to be a flexible tool in high-dimensional non-Gaussian dependence modeling and prediction for observational studies. This method consists of estimating the joint distribution
$F_{Y,\X}$ followed by inferences, such as point and interval predictions, based on the conditional distribution
$F_{Y|\X}$; $F_{Y,\X}$ is obtained from a vine pair-copula construction
for the dependence, combined with univariate marginal distributions
$F_{X_1},\ldots,F_{X_p},F_Y$; see
\cite{kraus2017d} and \cite{schallhorn2017d} for continuous response
and \cite{chang2019prediction} for responses that can be continuous or ordinal.

In this paper, we extend the methodology in \cite{chang2019prediction} for point and
interval predictions with a censored response variable and a set of discrete or continuous explanatory variables. Compared with parametric time-to-event analysis methods such as the Weibull Accelerated Failure Time (AFT) model, vine copulas separate the estimation of univariate marginal distributions from the modeling of multivariate dependence structure and
allow for more flexible forms for conditional quantile functions
$F^{-1}_{Y|\X}(p|\x)$ for $0<p<1$. For example, they could
be flattening in the extremes of the predictor space, something that
cannot be achieved with a linear method such as AFT.

The remaining sections are organized as follows. A brief introduction to time-to-event analysis is included in Section~\ref{sec: survival}. Section~\ref{sec: vine_cop} gives an overview of copula models and vine copulas. Our proposed method of using vine copulas for time-to-event prediction is presented in Section~\ref{sec: survival_copula}. Comparison criteria to assess time-to-event prediction performances
are summarized in Section~\ref{sec: eval}. Simulation studies comparing different methods are performed in Section~\ref{sec: simulation}. Section~\ref{sec: app} applies our proposed method to the analysis of the primary biliary cirrhosis (PBC) dataset. Finally, Section~\ref{sec: conclude} consists of concluding discussions.

\section{Survival or Time to Event Analysis} \label{sec: survival}

In this section, we first give a summary of regression methods for
a right-censored survival or time-to-event response. Comprehensive references for linear methods are
in \cite{lawless2011statistical} and \cite{klein2003survival}.


\subsection{Notation for Survival Analysis} \label{subsec: survival}


Suppose there is a random sample of size $n$ and the explanatory
variables are coded as $x_1,\ldots,x_p$.
For subject $i$, let $\x_i=(x_{i1}, \dots, x_{ip})^T$,
$y_i$ be the survival time or time to an event, and let $\delta_i$ be a binary event indicator
with value 1 if $y_i$ is observed and value 0 if $y_i$ is right-censored.
Let $F_Y$ be the distribution of $Y$ with density $f_Y$. 
The survival function is $S=1-F_Y$,
the cumulative hazard is $H=\log S$ and the hazard (rate) is $h=H'=f_Y/S$.

\subsection{Cox Proportional Hazards Model} \label{subsec: cox}

The hazard function for the Cox proportional hazards model has the form
\begin{equation}
h(y|\x_i) = h_0(y)\exp(\x_i^T\bbf), \label{eq: cox}
\end{equation}
where the baseline hazard $h_0(y)$ is obtained when $\x={\bf 0}$ and $\bbf = (\beta_1, \dots, \beta_p)^T$ is the coefficient vector.
The model is semi-parametric if the baseline hazard function $h_0(y)$ is unspecified and not estimated when optimizing the partial log-likelihood function. 
The survival function is
\begin{equation}
S(y|\x_i) = \exp\left(-H_0(y)\exp(\x_i^T\bbf)\right), \label{eq: cox_S1}
\end{equation}
where $H_0(y) = \int_{0}^{y}h_0(u)\mathrm{d}u$.
The Breslow estimator (\cite{breslow1972contribution}) is commonly used to estimate $H_0(y)$; it is a piecewise constant function that is undefined beyond the largest uncensored observation. The Cox model with this estimator for $H_0$ cannot be used to get reliable prediction intervals.
However, if a parametric form is used for $H_0$, and the estimation of
$H_0$ follows the maximum partial likelihood estimator of $\bbf$,
then prediction intervals can be obtained.


\subsection{Weibull Regression or Accelerated Failure Time Model} \label{subsec: aft}

The Accelerated Failure Time (AFT) model has stochastic representation 
for the log time-to-event:
  $$ \log y_i = \gamma_0 + \x_i^T\gbf + \sigma W_i, $$
where $\gbf = (\gamma_1, \dots, \gamma_p)^T$ is the coefficient vector,
$\gamma_0$ is the intercept, $\sigma>0$ is the scale parameter, $W_i$ are independent and identically distributed random variables with distribution $F_W$.
This is a location-scale model based on $F_W$ with location parameter $\gamma_0+\x_i^T\gbf$ that depends linearly on $\x_i$.
When $W_i$ has a standard extreme value distribution (for minima), $Y_i$ follows a Weibull distribution with survival function
\begin{equation}
S(y|\x_i) = \exp\Bigl[-\bigl(y\exp(-\gamma_0-\x_i^T\gbf)\bigl)^{1/\sigma}\Bigr]. \label{eq: aft}
\end{equation}
If we let $H_0(y) = (y/e^{\gamma_0})^{1/\sigma}$ and $\bbf =-\gbf/\sigma$ in Equation~\eqref{eq: cox_S1}, then it becomes the same as Equation~\eqref{eq: aft}. Therefore, the AFT model can be seen as a special parametric form of the Cox proportional hazards model with a specific baseline hazard function. 
For $H_0$ to be interpreted as a baseline hazard,
all continuous explanatory variables should be centered around 0 and all binary or ordinal discrete explanatory variables take value 0 for the baseline category.

The Weibull model is the most widely used parametric distribution for time-to-event analysis. 
Standard numerical maximum likelihood estimation techniques can be used to estimate the parameters. Point and interval predictions can be obtained from 
$F_{Y|\X}(\cdot|\x)=1-S(\cdot|\x)$ and $F^{-1}_{Y|\X}(p|\x)$ for $0<p<1$.

\subsection{Multivariate Distribution for Observational Study}

If $\X$ and $Y$ are obtained from a random sample, one flexible approach
is to estimate the joint distribution $F_{\X,Y}$ from which to get the
conditional distribution $F_{Y|\X}(\cdot|\x)$ for inferences such as
conditional quantiles. The case of a binary treatment variable is covered
if subjects are randomized to the treatment level. 
The vine pair-copula construction, which can handle a mix of discrete
and continuous variables, is an approach that can cover a wide range of
dependence structures and joint tail behavior. This construction is summarized
in the next section.

\section{Vine Copula Models} \label{sec: vine_cop}

In this section, vine copulas and notation for them are summarized for results in subsections.
For fitting a vine copula, we assume that
the dataset consists of $(x_{i1}, \dots, x_{id})$ for $i=1, \dots, n$; these vectors are considered as independent realizations of a continuous random vector $(X_1, \dots, X_d)$. The response variable can be considered as the last variable $X_d$.
After fitting univariate distributions to individuals variables,
probability integral transforms are applied to convert them to transformed
observation vectors in $[0,1]^d$.

\subsection{Introduction to Copulas} \label{subsec: copula}

A copula is a multivariate distribution function with univariate Uniform (0,1) margins. Sklar's theorem (\cite{sklar1959fonctions}) implies that a joint distribution $F$ is a composition of a copula $C$ and its univariate marginal distributions $F_1,\ldots,F_d$.
The copula associated with a $d$-variate distribution $F$ is a distribution function $C: [0,1]^d\to[0,1]$ with $U(0,1)$ margins that satisfies 
$ F(\x)=C(F_1(x_1),\dots,F_d(x_d))$, for $\x\in\mathbb{R}^d
$.
If $F$ is a continuous $d$-variate distribution function with univariate margins $F_1, \dots, F_d$, and quantile functions $F_1^{-1},\dots,F_d^{-1}$, then the copula
$C(\u)=F(F_1^{-1}(u_1),\dots,F^{-1}(u_d))$, for $\u\in[0,1]^d$,
is the unique choice. 
Vine copulas or pair-copula constructions are constructed from a sequence of bivariate copulas.
Detailed introductions to multivariate copula constructions can be found in \cite{joe2014dependence} 
and \cite{czado2019analyzing}.




\subsection{Overview of Vine Structure} \label{subsec: vine}

A regular vine (R-vine) is a nested set of trees where the edges in the first tree are the nodes of the second tree, the edges of the second tree are the nodes of the third tree, etc. Vines are useful in specifying the dependence structure for general multivariate distributions on $d$ variables with edges in the first tree representing
pairwise dependence and edges in subsequent trees representing conditional dependence.

The first tree in a vine represents $d$ variables as nodes and the bivariate dependence of $d-1$ pairs of variables as edges. The second tree describes the conditional dependence of $d-2$ pairs of variables conditioned on another variable; nodes are the edges in tree 1, and a pair of nodes can be connected if
there is a common variable in the pair. The third tree describes the conditional dependence of $d-3$ pairs of variables conditioned on two other variables; nodes are the edges in tree 2, and a pair of nodes could be connected if there are two common conditioning variables in the pair. This continues until tree $d-1$ has only one edge that describes the conditional dependence of two variables conditioned on the
remaining $d-2$ variables.

The definition of a vine as a sequence of trees is first given in \cite{bedford2002vines}. 
\begin{definition}
	\label{def: vine}
	$V$ is a vine on $d$ variables, with $E(V) = \bigcup_{\ell = 1}^{d-1} E(T_\ell)$ denoting the set of edges of $V$ if
	\setlist{nolistsep}
	\begin{enumerate}[noitemsep]
		\item
		$V = (T_1, \ldots, T_{d-1})$; 
		
		\item
		$T_1$ is a tree with nodes $N(T_1) = \{1, 2, \ldots, d\}$, and edges $E(T_1)$. For $\ell > 1$,
		$T_{\ell}$
		is a tree with nodes $N(T_{\ell}) = E(T_{\ell - 1})$;
		
		\item
		(proximity condition) For $\ell = 2, \ldots, d -1$, for $\{n_1, n_2\} \in E(T_{\ell})$, $\# (n_1 \triangle n_2) = 2$, where $\triangle$ denotes symmetric difference and $\#$ denotes cardinality.
		
	\end{enumerate}
\end{definition}

An R-vine can be represented by the edge sets at each level $E(T_\ell)$ or by a graph.
A vine array is a compact method to encode the conditional dependence in a vine. A vine array $A = (a_{\ell j})_{\ell = 1, \dots, d; j = \ell, \dots, d}$, for an R-vine $V = (T_1, \dots, T_{d-1})$ on $d$ elements is a $d\times d$ upper triangular matrix. It satisfies the following two conditions:
\begin{itemize}
	\itemsep=0pt 
	\item The diagonal of $A$ is a permutation of $(1, \dots, d)$.
	\item For $j = 2, \dots, d$, the $j$th column has $(a_{1j}, \dots, a_{j-1,j})$ being a permutation of $(a_{11}, \dots, a_{j-1,j-1})$. In the first row, $a_{1j}$ can be any element in $\{a_{11}, \dots , a_{j-1,j-1}\}$. For $\ell = 2, \dots, j-1$, the set $\{a_{1j}, \dots, a_{\ell j}\}$ is equal to $\{a_{1k}, \dots, a_{\ell - 1, k}, a_{kk}\}$ for at least one $k$ in columns $\ell, \dots, j - 1$.
\end{itemize}
For $\ell=2,\ldots,d-1$ and $\ell<j\le d$,
row $\ell$ and column $j$ of the vine array indicates that the variable $a_{\ell j}$
is connected to the variable $a_{jj}$ in tree $T_\ell$, conditioned on variables $a_{1j}, \dots, a_{\ell -1,j}$. In other words, the first $\ell$ rows of $A$ and the diagonal elements encode the $\ell$th tree $T_\ell$, such that
the edge  $[a_{\ell j}, a_{jj}|a_{1j}, \dots , a_{\ell-1,j}] \in E(T_\ell)$ summarizes the conditional dependence on $\ell-1$ variables, for $\ell + 1 \leq j \leq d$. 


To get a multivariate distribution from a vine, 
bivariate distributions are assigned to edges on the first tree and bivariate conditional distributions are assigned to edges on the subsequent trees. 
The bivariate copula corresponding to $a_{\ell j}$
of the vine array $A$ for $1 \leq l < j \leq d$ is denoted by $C_{a_{\ell j}, a_{jj}; S_{\ell j}}$ with density function $c_{a_{\ell j}, a_{jj}; S_{\ell j}}$, where $S_{\ell j} = \{a_{1j},\dots, a_{\ell-1,j}\}$ is the conditioning set for this position. Note that $S_{\ell j} = \emptyset$ if $\ell = 1$ and in tree 1,
$C_{a_{1j}, a_{jj}}$ summarizes the dependence of a pair of variables for $j=2,\ldots,d$. 
In tree $\ell\in \{2,\ldots,d-1\}$, the bivariate copula $C_{a_{\ell j}, a_{jj}; S_{\ell j}}$ is used to link the conditional distributions $F_{a_{\ell j}|S_{\ell j}}$ and $F_{a_{jj}|S_{\ell j}}$, 
and it summarizes the conditional dependence of variables indexed
as $a_{\ell j}$ and $a_{jj}$ given the variables in the index set $S_{\ell j}$.

Let $f_1, \dots, f_d$ be the univariate densities.
For absolutely continuous variables, the joint density of $(X_1, \dots , X_d)$ based on the vine structure specified by a vine array $A = (a_{\ell j})_{\ell = 1, \dots, d; j = \ell, \dots, d}$ can be decomposed as
\begin{equation}
	f_{1:d}(x_1, \dots, x_d) = \prod_{i=1}^{d}f_i(x_i) \cdot \prod_{\ell = 1}^{d-1} \prod_{j = \ell + 1}^{d}c_{a_{\ell j}, a_{jj}; S_{\ell j}}\left[F_{a_{\ell j}|S_{\ell j}}\left(x_{a_{\ell j}}|\x_{S_{\ell j}}\right), F_{a_{jj}|S_{\ell j}}\left(x_{a_{jj}}|\x_{S_{\ell j}}\right)\right], \label{eq: vine_density}
\end{equation}
where the conditional distributions $F_{a_{\ell j}|S_{\ell j}}$ and $F_{a_{jj}|S_{\ell j}}$ are determined in a recursive manner, using bivariate copulas on the edges in previous trees. Derivations of Equation~\eqref{eq: vine_density} are given in \cite{bedford2001probability},
Section 3.9 of \cite{joe2014dependence} and Theorem 5.15 in \cite{czado2019analyzing}.
See Section 3.9.5 of \cite{joe2014dependence} for a similar joint
density when some of the variables are discrete.

Since a joint distribution can be decomposed into univariate marginal distributions and a dependence structure among variables, estimation can proceed in a two-stage manner. The first step estimates the univariate marginal distributions $\widehat{F}_j$ for $j = 1, \dots, d$. The u-scores vectors 
$\u=(u_{i1},\ldots,u_{id})$ and $\u^-=(u^-_{i1},\ldots,u^-_{id})$ for $i=1, \dots, n$ 
are obtained by applying the probability integral transform: 
\begin{equation}
 u_{i,j} = \widehat{F}_j(x_{i,j}),
  \quad  u_{i,j}^- = \widehat{F}_j(x^-_{i,j}), \quad j=1,\ldots,p, 
  \ i=1,\ldots,n \label{eq: PIT} 
\end{equation}
The $u_{i,j}^-$ are needed only for discrete variables.
The second step fits a vine copula model based on the u-score vectors. 
In our approach, the second step involves two components: finding a vine structure describing the underlying dependence and deciding on suitable bivariate
parametric copula families to use on the edges of the vine. 

\section{Predicting Times to Event with Vine Copulas} \label{sec: survival_copula}

Assume the dataset contains $p$ explanatory variables $X_1, \dots, X_p$ and a response variable $Y$ as a sample of size $n$. The observed data are $(x_{i1}, \dots, x_{ip}, t_i)$ and the event indicator $\delta_i$ for $i = 1, \dots, n$. The observed time-to-event $t_i$ is equal to $y_i$ if $\delta_i = 1$. It is equal to $y_i^*$ where $y_i^* \leq y_i$ and $y_i$ is unobserved if $\delta_i = 0$. The data $(x_{i1}, \dots, x_{ip}, y_i)$ are considered as independent realizations of a random vector $(X_1, \dots, X_p, Y)$. To simplify the notation, we use $X_d$ to represent the response variable $Y$, i.e., the observed data are $(x_{i1}, x_{i2}, \dots, x_{id}) = (x_{i1}, \dots, x_{ip}, t_i)$ with $d = p + 1$.

Our goal is to extend the methodology in \cite{chang2019prediction}
to a right-censored response variable. The first step is to
estimate a van der Waerden correlation matrix; this is reasonable if a
Gaussian copula model is a first order model and variables are
monotonically related. For this step, Section \ref{subsec: correlation}
has the theory on computing a van der Waerden correlation when
one variable is right-censored and the other is continuous or discrete.
The second step is to obtain a vine structure such
that the response variable $Y$ is a leaf variable of each vine tree,
and pairs of variables with the strongest dependence are represented
in edges of low-order trees.


In the third step, we select the bivariate copula models on each edge in the selected vine structure. Copula models only associated with explanatory variables can be chosen based on existing algorithms. Copula models associated with the censored response variable can be chosen based on a stepwise procedure proposed in Section~\ref{subsubsec: selection_y}.
We also include an algorithm to evaluate the log-likelihood of a vine copula containing mixed discrete-continuous explanatory variables and a censored response variable in Section~\ref{subsubsec: llk_eval}. In the fourth step, point and interval predictions are obtained based on the fitted vine copula model. These steps are explained in details in the remainder of this section.

\subsection{Correlation Matrix Estimation} \label{subsec: correlation}



In this section, we propose a method to estimate the van der Waerden correlation between a continuous or discrete explanatory variable with a right-censored response variable when they are monotonically correlated. The main idea is to transform both variables to normal scores and then find the maximum likelihood estimator of the correlation coefficient assuming a bivariate Gaussian copula. The first $d-1$ rows and columns of the correlation matrix $\boldsymbol{R}_{d\times d}$ can be obtained by computing the usual correlation coefficients while the last row and column associated with the censored response can be obtained using the
results in the subsections given below.

\subsubsection{Van der Waerden Correlation for Continuous and Right-censored}

First we consider the case where $X=X_j$ is a continuous explanatory variable with observations $x_{1}, \dots, x_{n}$ and $Y$ is the censored continuous response with observations $(t_1,\delta_1), \dots, (t_n,\delta_n)$. The van der Waerden correlation between $X$ and $Y$ can be computed as follows. 
\begin{enumerate}
\item Let $\widehat{F}_{X}$ be the empirical CDF of $X$. Convert $\{x_i\}$ to u-scores with $u_{i1} = [\widehat{F}_{X}(x_{i}) + \widehat{F}_{X}(x_{i}^-)] / 2$ for $i = 1,\dots,n$, equivalently $u_{i1} = [\text{rank}\left(x_{i}) - 1/2\right] / n$. Further convert the u-scores to normal scores with $z_{i1} = \Phi^{-1}(u_{i1})$, where $\Phi^{-1}(\cdot)$ is the  inverse of the CDF of the standard normal distribution.

\item Based on $(t_i, \delta_i)$, obtain the Kaplan-Meier estimators for $\widehat{F}_Y(t_i)$ for $i = 1, \dots, n$. The Kaplan-Meier estimators are computed as
\[
\widehat{F}_Y(y) = 1 - \prod_{i: t_i<y}\bigl(1-d_i/n_i\bigr),
\]
where $t_i$ is a time when at least one event happens, $d_i$ is the number of events that happen at time $t_i$, and $n_i$ are the number of subjects known to have survived up to time $t_i$.

\item If $\delta_i = 1$, convert $t_i$ to u-score with $u_{i2} = [\widehat{F}_Y(t_i) + \widehat{F}_Y(t_i^-)] / 2$. Further convert the u-scores to normal scores with $z_{i2} = \Phi^{-1}(u_{i2})$. If $\delta_i = 0$, apply the same conversions, but the computed $z_{i2}$ is right-censored.

\item The van der Waerden correlation between $X$ and $Y$ is the maximum likelihood estimate of the log-likelihood function based on $(z_{i1}, z_{i2}, \delta_i)$ assuming the normal scores follow a bivariate Gaussian distribution:
\[
\widehat{\rho} = \argmax_{\rho\in(-1,1)}\log \mathcal{L}(\rho) = \argmax_{\rho\in(-1,1)} \sum_{i:\delta_i = 1}\log\phi_2(z_{i1}, z_{i2}; \rho) + \sum_{i:\delta_i = 0}\log\overline{\Phi}\left[(z_{i2} - \rho z_{i1})/(1-\rho^2)^{1/2}\right],
\]
where $\phi_2(\cdot, \cdot; \rho)$ is the bivariate Gaussian density with correlation $\rho$.
\end{enumerate}

\subsubsection{Van der Waerden Correlation for Discrete and Right-censored}

Next we consider the case where $X=X_j$ is a discrete explanatory variable with observations $x_{1}, \dots, x_{n}$ and $Y$ is the censored continuous response with observations $(t_1,\delta-1), \dots, (t_n,\delta_n)$. The van der Waerden correlation between $X$ and $Y$ can be computed as follows.
\begin{enumerate}
\item Let $\widehat{F}_{X}$ be the empirical CDF of $X$. Define $z_{i1l} = \Phi^{-1}(\widehat{F}_{X}(x_{i}^-))$ and $z_{i1u} = \Phi^{-1}(\widehat{F}_{X}(x_{i}))$. Then the z-score of $x_{i}$ satisfies $z_{i1} \in (z_{i1l}, z_{i1u}]$.

\item The observations $(t_i, \delta_i)$ for $Y$ can be computed to $z_{i2}$ following the same steps stated in the previous subsection.

\item The van der Waerden correlation between $X$ and $Y$ is the maximum likelihood estimate of the log-likelihood function based on $(z_{i1l}, z_{i1u} z_{i2}, \delta_i)$ assuming the normal scores follow a bivariate Gaussian distribution:
\begin{align*}
\widehat{\rho} &= \argmax_{\rho\in(-1,1)}\log \mathcal{L}(\rho)\\ 
&= \argmax_{\rho\in(-1,1)} \sum_{i:\delta_i = 1}\log\left\{\Phi\left[(z_{i1u} - \rho z_{i2})/(1-\rho^2)^{1/2}\right] - \Phi\left[(z_{i1l} - \rho z_{i2})/(1-\rho^2)^{1/2}\right]
\right\} \\
&\quad + \sum_{i:\delta_i = 0}\log\int_{z_{i1l}}^{z_{i1u}}\int_{z_{i2}}^{\infty}\phi_2(u,v;\rho)\mathrm{d}u\mathrm{d}v.
\end{align*}
\end{enumerate}

\subsection{Univariate Models} \label{subsec: univariate}

Before a vine copula is fitted for the dependence of $(\X,Y)$,
suitable univariate families should be chosen for each variable.
For continuous variables, there are many 2-parameter to 4-parameter 
families that can be considered, and maximum likelihood estimates can
be obtained. For a discrete or ordinal variable, the empirical
distribution could be used (perhaps after combining categories with low
counts). Candidate models can be compared via the Akaike information criterion (AIC) or Bayesian information criterion (BIC), and assessed for adequacy of fits (especially the tails)
via quantile-quantile (Q-Q) plots.


When fitting a univariate model $f(\cdot;\thbf)$, let $\thatbf$ be the 
maximum likelihood estimate. For a continuous variable $z$, 
When there is no censoring, let $z_{\{i\}}$
be the $i$th order value. The Q-Q plot consists of points
$(F^{-1}({i-0.5\over n};\thatbf), z_{\{i\}})$ for $i=1,\ldots,n$.

For a continuous variable $y$ that can be right-censored, 
let $y_{(1)}\le \cdots\le y_{(k)}$
be the observed uncensored values, and
let $p_i= [\Fhat_Y(y^+_{(i)})+\Fhat_Y(y^-_{(i)})]/2$ where
$\Fhat_Y$ is the Kaplan-Meier estimate of $F_Y$.
The Q-Q plot consists of points
$(F^{-1}(p_i;\thatbf), y_{(i)})$ for $i=1,\ldots,k$;
see Sections (3.3.1) and (3.3.4) of \cite{lawless2011statistical}.

\subsection{Bivariate Copula Selection} \label{subsec: selection}

After fitting the univariate marginal distributions and finding the vine structure $A_{d \times d}$, parametric bivariate copulas need to be selected and fitted for each edge in the vine from $T_1$ to $T_{d-1}$. This can be done in a two-step manner. In the first step, a vine copula model is fitted and bivariate copulas are selected for all the explanatory variables according to $A_{(d-1)\times(d-1)}$. These bivariate copulas are only associated with the fully observed explanatory variables and can be selected following standard procedures edge by edge sequentially. In the second step, the parameters of the bivariate copulas selected in the first step are fixed. However, due to censoring, bivariate copulas associated with the censored response variable cannot be estimated edge by edge using the standard method. We propose an algorithm to evaluate the log-likelihood of a vine copula model containing mixed discrete-continuous explanatory variables and a censored continuous response variable. Using this algorithm, these bivariate copulas can be selected in a stepwise procedure.

\subsubsection{Bivariate Copula Selection for Explanatory Variables} \label{subsubsec: selection_x}

For all the fully observed explanatory variables, a vine copula model can be fitted based on the vine structure $A_{(d-1)\times(d-1)}$. This is possible since the response variable is only contained in a leaf node in all the trees. As a result, $d$ is not contained in $A_{(d-1)\times(d-1)}$.

Depending on whether the two variables are discrete or continuous, the copula density function has different expressions. We define $u_j^+ = u_j = F_j(x_j)$ and $u_j^- = \lim\limits_{t\to x_j^-} F_j(t) = F_j(x_j^-)$. For continuous variables, $u_j^+ = u_j^-$. We further define a function $\ctil_{jk|S}(\F_{j|S}, \F_{k|S}; \thbf)$ to represent the copula density for a pair of CDFs $\F_{j|S} = (F_{j|S}^+, F_{j|S}^-)$ and $\F_{k|S} = (F_{k|S}^+, F_{k|S}^-)$. The conditioning set $S$ can be either empty or non-empty.
\begin{itemize}
	\item If both $X_j$ and $X_k$ are continuous, then
	\[
	\ctil_{jk|S}(\F_{j|S}, \F_{k|S}; \thbf) = c_{jk|S}(F_{j|S}, F_{k|S}; \thbf).
	\]
	\item If $X_j$ is discrete and $X_k$ is continuous, then 
	\[
	\ctil_{jk|S}(\F_{j|S}, \F_{k|S}; \thbf) = \frac{C_{j|k;S}\left(F_{j|S}^+\big|F_{k|S}; \thbf\right) - C_{j|k;S}\left(F_{j|S}^-\big|F_{k|S}; \thbf\right)}{F_{j|S}^+ - F_{j|S}^-}.
	\]
	\item If $X_j$ is continuous and $X_k$ is discrete, then
	\[
	\ctil_{jk|S}(\F_{j|S}, \F_{k|S}; \thbf) = \frac{C_{k|j;S}\left(F_{k|S}^+\big|F_{j|S}; \thbf\right) - C_{k|j;S}\left(F_{k|S}^-\big|F_{j|S}; \thbf\right)}{F_{k|S}^+ - F_{k|S}^-}.
	\]
	\item If both $X_j$ and $X_k$ are discrete, then
	\begin{align*}
	\ctil_{jk|S}(\F_{j|S}, \F_{k|S}; \thbf) &= \Big[C_{jk;S}\left(F_{j|S}^+, F_{k|S}^+; \thbf\right) - C_{jk;S}\left(F_{j|S}^-, F_{k|S}^+; \thbf\right) - C_{jk;S}\left(F_{j|S}^+, F_{k|S}^-; \thbf\right) \\
	&\quad + C_{jk;S}\left(F_{j|S}^-, F_{k|S}^-; \thbf\right)\Big] \Big/ \Big[
	\left(F_{j|S}^+ - F_{j|S}^-\right)\left(F_{k|S}^+ - F_{k|S}^-\right)\Big].
	\end{align*}
\end{itemize}

Suppose there are $M$ bivariate copula candidate families for each edge in the vine. In the first tree, consider the edge $[a_{1j},a_{jj}]$. The log-likelihood of the bivariate copula model $(m)$ on this edge is
\begin{equation}
\mathcal{L}_{a_{1j}, a_{jj}}\left(\thbf^{(m)}\right) = \sum_{i = 1}^{n} \log \ctil^{(m)}_{a_{1j}, a_{jj}}\bigl[\u_{i, a_{1j}}, \u_{i, a_{jj}}; \thbf^{(m)} \bigr], \label{eq: llk_first_level}
\end{equation}
where $\u_{ij}$ represents $(u_{ij}^+, u_{ij}^-)$. Commonly used model selection criteria include AIC and BIC.
For each candidate parametric bivariate copula model on an edge, the maximum likelihood parameter estimator $\widehat{\thbf}^{(m)}$ is obtained. The parametric copula model with the lowest AIC or BIC is selected for that edge. 

In tree $\ell\in \{2,\ldots,d-2\}$, consider the edge $[a_{\ell j},a_{jj};S_{\ell j}]$. 
Based on the fitted copulas at the previous levels, the pseudo observations $\Cbf_{a_{\ell j}|S_{\ell j}}\bigl(\u_{i, a_{\ell j}}|\u_{i, S_{\ell j}}\bigr) = \bigl(C_{a_{\ell j}|S_{\ell j}}^+, C_{a_{\ell j}|S_{\ell j}}^-\bigr)$ and $\Cbf_{a_{jj}|S_{\ell j}}\bigl(\u_{i, a_{jj}}|\u_{i, S_{\ell j}}\bigr) = \bigl(C_{a_{jj}|S_{\ell j}}^+, C_{a_{jj}|S_{\ell j}}^-\bigr)$ can be obtained.
The log-likelihood of the bivariate copula model $(m)$ on this edge is
\begin{equation}
\mathcal{L}_{a_{\ell j}, a_{jj}; S_{\ell j}}\left(\thbf^{(m)}\right) = \sum_{i = 1}^{n} \log \ctil^{(m)}_{a_{\ell j}, a_{jj}; S_{\ell j}}\left[\Cbf_{a_{\ell j}|S_{\ell j}}\left(\u_{i, a_{\ell j}}|\u_{i, S_{\ell j}}\right), \Cbf_{a_{jj}|S_{\ell j}}\left(\u_{i, a_{jj}}|\u_{i, S_{\ell j}}\right); \thbf^{(m)} \right], \label{eq: llk_other_levels}
\end{equation}
where $\u_{i, S_{\ell j}} = \{(u_{ik}^+, u_{ik}^-): k\in S_{\ell j}\}$. 
AIC and BIC can again be used as the selection criteria.

This approach to selecting bivariate copulas
is an extension of a method
implemented in the \verb|VineCopula| R package (\cite{schepsmeier2019vinecopula}), to accommodate a mix of continuous and discrete variables. 
Following this approach, bivariate copula models can be selected and fitted for the explanatory variables.

\subsubsection{Log-Likelihood Evaluation for Vine Copulas with Censored Response} \label{subsubsec: llk_eval}


In the vine, let $\pi_\ell$ be the index of the $x$-variable
linked to the response variable $y$ in tree $\ell$ for $\ell=1,\ldots,p$.
For the vine array $A$, $\pi_\ell$ is in row $\ell$ and column $d=p+1$.

For $y_i$ that is uncensored, the contribution to the vine copula
log-likelihood involves the copula density 
$c_{X_{\pi_\ell},Y; X_{\pi_1},\ldots,X_{\pi_{\ell-1}}}(a_i,b_i)$ where
$a_i,b_i$ are univariate CDF values for $\ell=1$ and are
conditional CDF values (recursively computed based on earlier trees)
for $\ell>1$.  For $y_i$ that is right-censored ($\delta_i=0$), the contribution of
the log-likelihood is more complicated as it involves that
conditional distribution $C_{Y|X_1,\ldots,X_p}$ and this 
involves the copulas 
\begin{equation}
C_{X_{\pi_\ell},Y; X_{\pi_1},\ldots,X_{\pi_{\ell-1}}} 
  \label{eq: Ycopulas}
\end{equation}
altogether for $\ell=1,\ldots,p$. 

This means that with parametric families for bivariate copulas,
the parameters in Equation~\eqref{eq: Ycopulas} cannot be estimated separately
and sequentially as in \cite{chang2019prediction}.
In this section, we indicate the algorithm for evaluating the log-likelihood of an R-vine copula model containing mixed discrete-continuous explanatory variables and a right-censored continuous response variable by modifying Algorithm 4 in Chapter 6 of \cite{joe2014dependence} and Algorithm 3.1 in \cite{chang2019prediction}.

Assume variables have permuted indices so that the vine array has
$1,\ldots,d$ on the diagonal.
The inputs to the algorithm are the vine array $A_{d\times d} = (a_{kj})$ with $a_{jj} = j$ for $j = 1, \dots, d$ on the diagonal, the bivariate copula family matrix $F_{d\times d}$ for all the edges in the vine, as well as two sets of u-score vectors $\u_i^+ = (u_{i1}^+, \dots, u_{id}^+)$ and $\u_i^- = (u_{i1}^-, \dots, u_{id}^-)$ for $i = 1, \dots, n$, where $u_{ij}^+ = F_j(x_{ij})$ and $u_{ij}^- = F_j(x_{ij}^-)$ for $1\leq j \leq d-1$. Note that for the censored response variable, we only need to consider $u_{id}^+$ and $C_{d|S_{\ell j}}^+\left(u_{id}^+|\u_{i, S_{\ell j}}\right)$ for the two terms $\u_{id}$ and $\Cbf_{d|S_{\ell j}}\left(\u_{id}|\u_{i, S_{\ell j}}\right)$ in Equations~\eqref{eq: llk_first_level} and \eqref{eq: llk_other_levels}, since the response variable is continuous. Note that the parameter vector $\thbf$ for each bivariate copula model is omitted in the algorithm for simplicity. 

When all the bivariate families are fixed, Algorithm~\ref{alg: llk} returns the value of the log-likelihood function of an R-vine involving a censored response variable.

\begin{algorithm}
\caption{Log-likelihood evaluation for the R-vine density with continuous or discrete predictors as variables $1, \dots, d-1$ and a censored response as variable $d$.}
\footnotesize
\begin{algorithmic}[1]
\State Compute $M = (m_{kj})$ in the upper triangle, where $m_{kj} = \max\{a_{1j}, \dots, a_{kj}\}$ for $k = 1, \dots, j-1$, $j = 2, \dots, d$.
\State Compute the $I = (I_{kj})$ indicator array as in Algorithm 5 in \cite{joe2014dependence}.
\State Set $s_{ij}^+ \gets u_{i,a_{1j}}^+$, $s_{ij}^- \gets u_{i,a_{1j}}^-$, $w_{ij}^+ \gets u_{ij}^+$, $w_{ij}^- \gets u_{ij}^-$, for $j = 1, \dots, d$. 
\State Set $\text{loglik} \gets \sum_{i:\delta_i = 1}\sum_{j = 2}^{d}\log \ctil_{a_{1j}, j}(\sbf_{ij},\w_{ij}) + \sum_{i:\delta_i = 0}\sum_{j = 2}^{d-1}\log \ctil_{a_{1j}, j}(\sbf_{ij},\w_{ij})$, where $\sbf_{ij} = (s_{ij}^+, s_{ij}^-)$, $\w_{ij} = (w_{ij}^+, w_{ij}^-)$, and the function $\ctil$ is defined earlier. This is the log-likelihood function at the first level.
\For{$l = 2, \dots, d-1$}
\For{$j = l, \dots, d$}
\If{$I_{l-1,j} = 1$}
\If{isDiscrete(variable $j$)}
\State Set $v_{ij}'^+ \gets [C_{a_{l-1,j},j;a_{1j},\dots,a_{l-2,j}}\left(s_{ij}^+, w_{ij}^+\right) - C_{a_{l-1,j},j;a_{1j},\dots,a_{l-2,j}}\left(s_{ij}^+, w_{ij}^-\right)]/(w_{ij}^+ - w_{ij}^-)$.
\State Set $v_{ij}'^- \gets [C_{a_{l-1,j},j;a_{1j},\dots,a_{l-2,j}}\left(s_{ij}^-, w_{ij}^+\right) - C_{a_{l-1,j},j;a_{1j},\dots,a_{l-2,j}}\left(s_{ij}^-, w_{ij}^-\right)]/(w_{ij}^+ - w_{ij}^-)$.
\Else 
\State Set $v_{ij}'^+ \gets C_{a_{l-1,j}|j;a_{1j},\dots,a_{l-2,j}}\left(s_{ij}^+\big| w_{ij}^+\right)$
and $v_{ij}'^- \gets C_{a_{l-1,j}|j;a_{1j},\dots,a_{l-2,j}}\left(s_{ij}^-\big| w_{ij}^+\right)$.
\EndIf
\EndIf
\If{isDiscrete(variable $a_{l-1, j}$)}
\State Set $v_{ij}^+ \gets [C_{a_{l-1,j},j;a_{1j},\dots,a_{l-2,j}}\left(s_{ij}^+, w_{ij}^+\right) - C_{a_{l-1,j},j;a_{1j},\dots,a_{l-2,j}}\left(s_{ij}^-, w_{ij}^+\right)]/(s_{ij}^+ - s_{ij}^-)$.
\State Set $v_{ij}^- \gets [C_{a_{l-1,j},j;a_{1j},\dots,a_{l-2,j}}\left(s_{ij}^+, w_{ij}^-\right) - C_{a_{l-1,j},j;a_{1j},\dots,a_{l-2,j}}\left(s_{ij}^-, w_{ij}^-\right)]/(s_{ij}^+ - s_{ij}^-)$.
\Else
\State Set $v_{ij}^+ \gets C_{j|a_{l-1,j};a_{1j},\dots,a_{l-2,j}}\left(w_{ij}^+\big| s_{ij}^+\right)$ and
$v_{ij}^- \gets C_{j|a_{l-1,j};a_{1j},\dots,a_{l-2,j}}\left(w_{ij}^-\big| s_{ij}^+\right)$.
\EndIf
\EndFor
\For{$j = l+1, \dots, d$}
\If{$a_{lj} = m_{lj}$}
\State Set $s_{ij}^+ \gets v_{i,m_{lj}}^+$ and $s_{ij}^- \gets v_{i,m_{lj}}^-$.
\Else 
\State Set $s_{ij}^+ \gets v_{i,m_{lj}}'^+$ and $s_{ij}^- \gets v_{i,m_{lj}}'^-$.
\EndIf
\State Set $w_{ij}^+ \gets v_{ij}^+$ and $w_{ij}^- \gets v_{ij}^-$.
\EndFor
\If{$l < d-1$}
\State Update $\text{loglik} \gets \text{loglik} + \sum_{i:\delta_i = 1}\sum_{j = l+1}^{d}\log \ctil_{a_{lj}, j;a_{1j},\dots,a_{l-1,j}}(\sbf_{ij},\w_{ij}) + \sum_{i:\delta_i = 0}\sum_{j = l+1}^{d-1}\log \ctil_{a_{lj}, j;a_{1j},\dots,a_{l-1,j}}(\sbf_{ij},\w_{ij})$.
\Else
\If{isDiscrete(variable $a_{d-1, d}$)}
\State  Update 
\begin{align*}
\text{loglik} \gets &\text{loglik} + \sum_{i:\delta_i = 1}\log \ctil_{a_{d-1,d}, d; a_{1d},\dots,a_{d-2,d}}(\sbf_{id}, \w_{id}) \\
&+\sum_{i:\delta_i = 0}\log\left[1-\frac{C_{a_{d-1,d},d;a_{1d},\dots,a_{l-2,d}}\left(s_{id}^+, w_{id}^+\right) - C_{a_{d-1,d},d;a_{1d},\dots,a_{l-2,j}}\left(s_{id}^-, w_{id}^+\right)}{s_{id}^+ - s_{id}^-}\right].
\end{align*}
\Else
\State Update $\text{loglik} \gets \text{loglik} + \sum_{i:\delta_i = 1}\log \ctil_{a_{d-1,d}, d; a_{1d},\dots,a_{d-2,d}}(\sbf_{id}, \w_{id}) + \sum_{i:\delta_i = 0}\log\left[ 1-C_{d|a_{d-1,d};a_{1d},\dots,a_{d-2,d}}(w_{id}^+| s_{id}^+)\right]$.
\EndIf
\EndIf
\EndFor
\State Return loglik.
\end{algorithmic} \label{alg: llk}
\end{algorithm}

\subsubsection{Bivariate Copula Selection for the Response Variable} \label{subsubsec: selection_y}
With the algorithm proposed in the previous section, standard maximum likelihood estimation techniques can be applied to estimate the copula parameters once the bivariate copula families are determined. In order to select the bivariate copula families associated with the censored response variable, we proposed a stepwise method in this section. The detailed steps are as follows.

\begin{enumerate}
	\item The bivariate copula families that are only associated with the explanatory variables are selected using the procedure in Section~\ref{subsubsec: selection_x}. The parameter estimates of these copulas are fixed when selecting the bivariate copulas associated with the censored response variable.
	\item Copulas for the edges associated with the response variable, i.e., edge $[a_{1d}, a_{dd}]$ on the first level and edges $[a_{\ell d}, a_{dd};S_{\ell d}]$ on the remaining levels for $\ell = 2, \dots, d-1$, are all set to bivariate Gaussian copula as a starting point.
	\item Starting from the first level, the bivariate copula for the edge $[a_{1d}, a_{dd}]$ is changed to one of the $M$ bivariate copula candidate families. The copula family that minimizes criteria such as AIC or BIC of the entire vine copula model with all the bivariate copulas associated with the response variable at higher levels set to Gaussian is chosen for the first level. This bivariate copula family is fixed but the corresponding parameters are re-estimated every time when choosing bivariate copulas at higher levels.
	\item For levels $\ell = 2, \dots, d-1$, repeat Step 3 to select the bivariate copula family for the edge $[a_{\ell d}, a_{dd};S_{\ell d}]$ with all the bivariate copulas associated with the response at higher levels set to Gaussian.
	\item At the last level, when the last bivariate copula family for the edge $[a_{d-1, d}, a_{dd};S_{d-1, d}]$ is chosen, parameters of all the bivariate copulas associated with the censored response are re-estimated to maximize the log-likelihood. 
\end{enumerate}

\subsection{Time to Event Prediction} \label{subsec: prediction}

Using the proposed algorithm and procedures in Section~\ref{subsec: selection}, a vine copula model can be fitted and estimated for the mixed discrete-continuous explanatory variables and the censored response variable. With the fitted vine copula model, it is possible to apply Algorithm 3.1 in \cite{chang2019prediction} to compute the conditional CDF of the response variable given the explanatory variables evaluated at $u$, i.e., 
 $$\pi(u|x_1,\dots,x_{d-1}) = \mathbb{P}\left(F_Y(Y) \leq u|x_1,\dots,x_{d-1}\right) = C_{d|1,\dots,d-1}(u|F_1(x_1),\dots,F_{d-1}(x_{d-1})).$$
Given a quantile $q$, the solution to the equation $\pi(u|x_1,\dots,x_{d-1}) = q$, i.e., $u = \pi^{-1}(q|x_1,\dots,x_{d-1})$, is the desired quantile. We can further apply the inverse probability integral transform to obtain the predicted event times based on the corresponding univariate model.

\section{Prediction Performance Evaluation} \label{sec: eval}

In this section, we discuss model comparisons and evaluations for predicting a censored response variable. First, we propose a multiple imputed version of the performance metrics to adapt them for a censored response. We then give an example where a vine copula model cannot be well approximated by the Cox or AFT model.

\subsection{Performance Evaluation} \label{subsec: metrics}

\subsubsection{Performance Metrics for Uncensored Response} \label{subsubsec: metrics_uncensored}

For an uncensored numeric response variable, prediction methods are usually compared based on mean absolute error of prediction, root mean squared error of prediction, and interval score of prediction intervals. A reference for interval score (IS) is \cite{gneiting2007strictly}.

Assume a training set is used to fit the models and a test set of size $n_{\text{test}}$ is used to compare the models. The definitions of these three measures are given below.

\begin{itemize}
\itemsep=0pt
\item The mean absolute error (MAE) and root mean squared error (RMSE) measure a model's performance on point estimations:
\begin{equation}
\text{MAE}(\mathcal{M}) =\frac{1}{n_{\text{test}}}\sum_{i=1}^{n_\text{test}}\left|\ytil_i-\widehat{\ytil}^{\mathcal{M}}_i\right|, \quad
\text{RMSE}(\mathcal{M}) = \sqrt{\frac{1}{n_{\text{test}}}\sum_{i=1}^{n_\text{test}}\left(\ytil_i-\widehat{\ytil}^{\mathcal{M}}_i\right)^2} \,, 
\label{eq: MAE} 
\end{equation}
where $\ytil_i$ is true value of the response variable of the $i$th sample in the test set and $\widehat{\ytil}^{\mathcal{M}}_i$ is the predicted conditional expectation or conditional median based on a fitted model $\mathcal{M}$.
	
	
\item The interval score (IS) is a scoring rule for quantile and interval predictions. In the case of the central $100(1-\alpha)$\% prediction interval, let $\widehat{l}^{\mathcal{M}}_i$ and $\widehat{u}^{\mathcal{M}}_i$ be the predicted quantiles at levels $\alpha/2$ and $1-\alpha/2$ by a fitted model $\mathcal{M}$ for the response variable of the $i$th sample in the test set. The interval score is defined as
\begin{equation}
\text{IS}(\mathcal{M}) = 
\frac{1}{n_{\text{test}}}\sum_{i=1}^{n_\text{test}}\Bigl[\bigl(\widehat{u}^{\mathcal{M}}_i-\widehat{l}^{\mathcal{M}}_i \bigr) + \frac{2}{\alpha}\bigl(\widehat{l}^{\mathcal{M}}_i - \ytil_i\bigr)\mathbb{I}\{\ytil_i < \widehat{l}^{\mathcal{M}}_i\} + \frac{2}{\alpha}\bigl(\ytil_i - \widehat{u}^{\mathcal{M}}_i\bigr)\mathbb{I}\{\ytil_i > \widehat{u}^{\mathcal{M}}_i\}\Bigr]. \label{eq: IS} 
\end{equation}
Smaller interval scores imply superior prediction performance
that has prediction intervals that are not too long and do not miss the
``true value'' by much when the prediction interval does not contain
the true value in the test set.
\end{itemize}

These metrics can be applied to evaluate and compare the performance of different prediction methods in simulation studies when the true values of the unobserved response variable are known for the censored cases. Nonetheless, for time-to-event datasets collected from real-life observational studies, the true values remain unknown when there is censoring. Therefore, alternative metrics are needed to evaluate the prediction performance for time-to-event data.

\subsubsection{Performance Metrics for Censored Response} \label{subsubsec: metrics_survival}

In time-to-event analysis, a common metric to evaluate the prediction performance of a model is the concordance index (C-index) introduced in \cite{harrell1982evaluating}; see also \cite{wang2019machine}. C-index compares the ordering of the predicted relative risks or survival probabilities of different subjects instead of event times. For each pair of subjects, it can be considered as a concordant pair, a discordant pair, or an indeterminable pair. C-index is defined as the number of concordant pairs divided by the total number of concordant and discordant pairs. Nevertheless, larger C-index values do not necessarily imply more accurate prediction results.

In this section, we adapt the performance metrics to right censoring by applying multiple imputation. In statistics, imputation is the process of replacing missing data with substituted values and the censored responses in time-to-event analysis can be considered as a form of missing data. Proposed by \cite{rubin2004multiple}, multiple imputation averages the outcomes across multiple imputed datasets to account for the increased noise due to single imputation. Given a measure for an uncensored response variable, its multiple imputed version for the censored response variable can be computed by the following steps:
\begin{enumerate}
\item \textit{Imputation}: Imputed values are drawn from a fitted conditional distribution of $F_{Y|\X}$ satisfying the condition that the imputed values should be greater than the censored response $y_i^*$. This can be done by first drawing a random sample $U$ from the Uniform $[F_{Y|\X}(y_i^*|\x_i), 1]$ distribution and then applying the inverse of the probability integral transform to obtain the imputed value $y^{\text{mi}} = F^{-1}_{Y|\X}(U|\x_i)$. Repeat this procedure $m$ times. At the end of this step, there should be $m$ imputed values.

\item \textit{Computation}: Each of the $m$ imputed values is considered as the true value of the response variable $\ytil_i$. For each subject, if it is censored, the evaluation metrics are computed based on the predicted and imputed values. If it is fully observed, the evaluation metrics are computed based on the predicted and true values. At the end of this step, there should be $m$ sets of evaluation metrics.

\item \textit{Pooling}: The $m$ sets of evaluation metrics are consolidated into one by taking the mean.
\end{enumerate}
The multiple imputed version of the evaluation metrics can be computed as an average of the two components: the raw metrics when the response variable is fully observed and the multiple imputed metrics when the response variable is censored. For example, the multiple imputed MAE can be defined as
\[
\text{MAE}^{\text{mi}}(\mathcal{M}) =\frac{1}{n_{\text{test}}}\sum_{i=1}^{n_\text{test}}\left\{\mathbb{I}(\delta_i = 1)\left|\ytil_i-\widehat{\ytil}^{\mathcal{M}}_i\right| + \mathbb{I}(\delta_i = 0)\frac{1}{m}\sum_{k=1}^{m}\left|y^{\text{mi}}_{ik}-\widehat{\ytil}^{\mathcal{M}}_i\right|\right\}.
\]
The multiple imputed RMSE and IS can be defined in a similar manner.

When the number of imputations $m$ is large and the fitted conditional distribution $F_{Y|\X}$ is close to the true distribution, the multiple imputed version of an evaluation metric is a good approximation of the true metric.

\subsection{Comparing Copula-based Prediction to Other Models} \label{subsec: compare_vine}

Copulas are often used to model high-dimensional dependence structures
with tail properties that cannot be accommodated by other models.
Conditional quantile functions for the response variable given other
observed variables can have a variety of shapes including 
monotonically increasing or asymptotically flat in the tails.
In this section, we provide some examples where conditional distributions from copula-based models
can or cannot be approximated by the Cox proportional hazards ratio or AFT assumptions.

Based on the hazard function for the Cox proportional hazards model in Equation~\eqref{eq: cox}, we know that when the explanatory variables $X_1, \dots, X_{j-1}, X_{j+1}, \dots, X_{p}$ are fixed, at any value of $Y = y$, the hazard function $h(y|\X = \x)$ is either monotonically increasing or decreasing with respect to the $x_j$ depending on the signs of $\beta_j$. Therefore, the cumulative hazard function
$ H(y|\x) = \int_{0}^{y}h(t|\x)\,\mathrm{d}t$
should also be either monotonically increasing or decreasing with respect to $x_j$ at any value of $y$ for any $j$ under the proportional hazards assumption.

Similarly, the AFT model assumes the survival function in the form of Equation~\eqref{eq: aft}. Hence, the cumulative hazard function is
\[
H(y| \x) = -\log S(y|\x) = \left[y\exp(-\gamma_0-\x^T\gbf)\right]^{1/\sigma}.
\]
Again, depending on $\gamma_j$ and $\sigma$, when $x_1, \dots, x_{j-1}, x_{j+1}, \dots, x_{p}$ are fixed, at any value of $Y = y$, $H(y|\x)$ is either monotonically increasing or decreasing with respect to $x_j$ for any $j$.

To compare copula models with the Cox and AFT models, we first consider the case where the explanatory variables $X_1, \dots, X_p$ and the response variable $Y$ marginally follow univariate Gaussian distributions and the joint distribution of their u-scores is characterized by a Gaussian copula. This is equivalent to the case where the explanatory variables and the response variable follow a multivariate Gaussian distribution. Without loss of generality, assume that $(X_1, \dots, X_p, Y) \sim N(0, \Sbf)$. Consider partitioning $\Sbf$ into
\[
\Sbf = \begin{bmatrix}
\Sbf_{XX} & \Sbf_{XY} \\
\Sbf_{YX} & \Sigma_{YY} \\
\end{bmatrix},
\]
where $\Sbf_{XX}$ is the $p\times p$ covariance matrix of the explanatory variables $X_1, \dots, X_p$ and $\Sigma_{YY}$ is the variance of the response variable $Y$. The conditional distribution of the response variable given the explanatory variables is $Y|(\X = \x) \sim N(\Sbf_{XY}\Sigma_{YY}^{-1}\x, \Sbf_{XX} - \Sbf_{XY}\Sigma_{YY}^{-1}\Sbf_{YX})$. The covariance matrix of this conditional Gaussian distribution is a constant while the mean of this conditional Gaussian distribution is a linear function of the explanatory variables. As a result, when $x_1, \dots, x_{j-1}, x_{j+1}, \dots, x_p$ are fixed, at any value of $Y = y$, the mean of this conditional Gaussian distribution is either monotonically increasing or decreasing with respect to the $x_j$ for any $j$. Therefore, the survival probability $S(y|\x)$ at any value of $Y = y$ is also monotonically increasing or decreasing with respect to the $x_j$ since the mean of the Gaussian distribution is monotonically changing while the variance is a constant. It further implies that $H(y|\X = \x)$ is also monotonically changing with respect to $x_j$. This shows that the Cox proportional hazards or AFT assumptions can be approximated by Gaussian copula models. When the simplifying assumption for conditional copulas is satisfied, a vine copula model with bivariate Gaussian copulas on all the edges is a good approximation of a multivariate Gaussian copula model, thus can approximate the Cox proportional hazards or AFT assumptions well.

In contrast, when a vine copula model has non-Gaussian bivariate copulas on the edges, it can be easily shown by simulation studies that the Cox proportional hazards or AFT assumptions are violated. For example, we consider the vine copula model (A) as follows:

\noindent\textit{Model (A). 5 dimensional C-vine copula model}: 

Variables 1--4 are continuous explanatory variables and variable 5 is the response variable. Each explanatory variable follows a univariate norm\textbf{}al $(0,1)$ distribution. The response variable follows an Exponential (1) distribution. The vine array, bivariate copula family array, and bivariate copula parameter array of this vine copula model are as follows:
\[
A = 
\begin{bmatrix}
1 & 1 & 1 & 1 & 1\\
& 2 & 2 & 2 & 2\\
&   & 3 & 3 & 3\\
&   &   & 4 & 4\\
&   &   &   & 5\\
\end{bmatrix},
F =
\begin{bmatrix}
- & \text{G} & \text{G} & \text{G} & \text{G}\\
& - & \text{t} & \text{t} & \text{t}\\
& & - & \text{F} & \text{F}\\
& & & - & \text{N}\\
& & & & - \\
\end{bmatrix},
P =
\begin{bmatrix}
- & 4.67 & 2.67 & 5.32 & 3.09\\
& - & 0.75 (6) & 0.61 (6) & 0.7 (6)\\
& & - & 3.14 & 3.86\\
& & & - & 0.25\\
& & & & - \\
\end{bmatrix},
\]
where N, t, F, and G in the copula family matrix stands for bivariate Gaussian (normal), t, Frank, and Gumbel copulas, respectively. The t copula family has two parameters while all the other copula families have one parameter. 

\begin{figure}[!ht] 
	\centering
	\includegraphics[width=\linewidth]{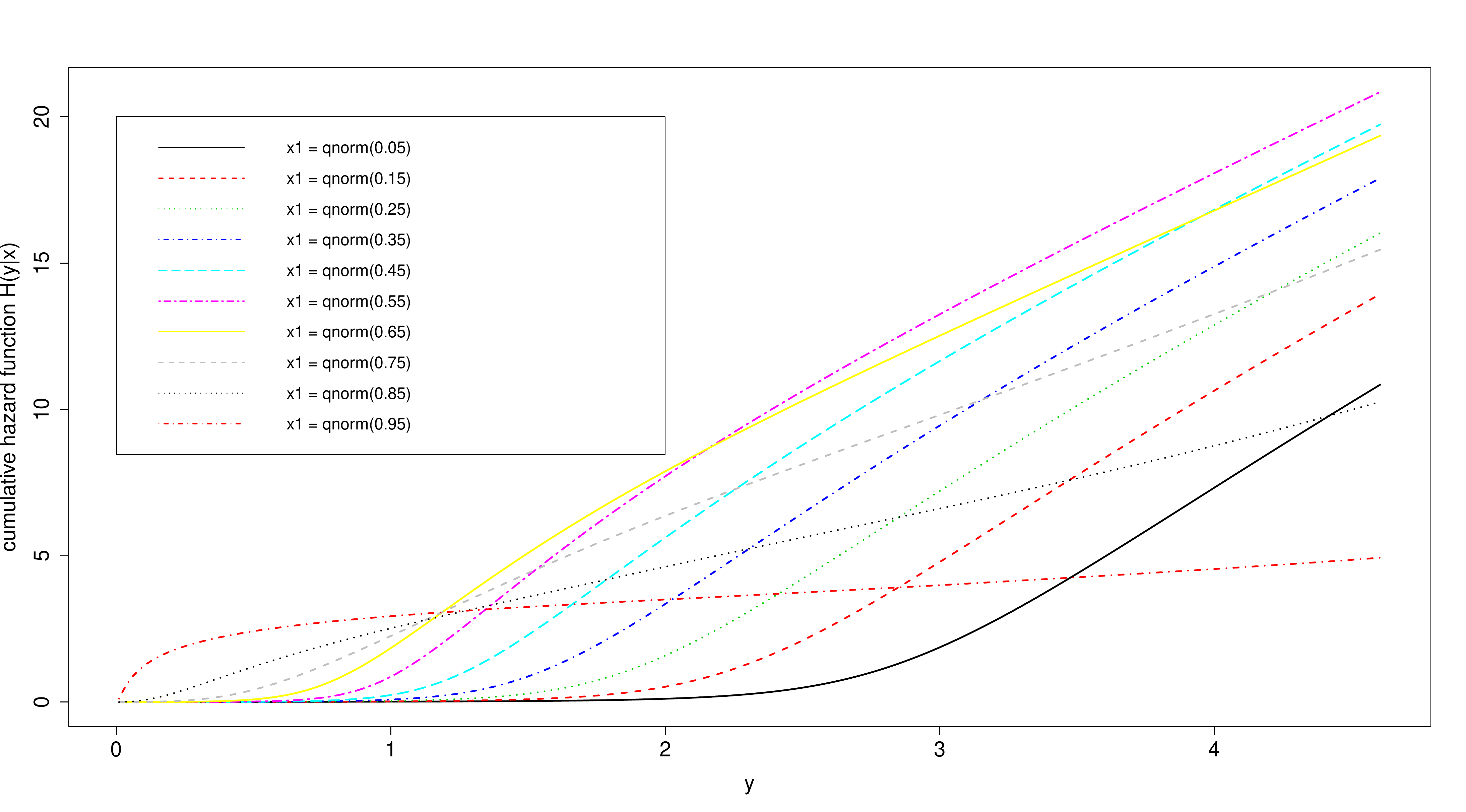}
	\caption{Plot of the cumulative hazard function $H(y|x_1, x_2, x_3, x_4)$ against $y$ at different values of $x_1$ from $\Phi^{-1}(0.05)$ to $\Phi^{-1}(0.95)$ with a randomly generated set of explanatory variables $X_2$, $X_3$, and $X_4$.}
	\label{fig: ph_check}
\end{figure}

With a randomly generated set of explanatory variables $X_2$, $X_3$, and $X_4$, the plot of the cumulative hazard function $H(y|x_1, x_2, x_3, x_4)$ against $y$ at different values of $x_1$ from $\Phi^{-1}(0.05)$ to $\Phi^{-1}(0.95)$ is shown in Figure~\ref{fig: ph_check}. If the proportional hazards or AFT assumption holds, the cumulative hazard function should be monotonically increasing with $X_1$ given any value of $y$. In other words, these lines should be ordered from $\Phi^{-1}(0.05)$ at the bottom to $\Phi^{-1}(0.95)$ at the top for all the $y$ values. However, it can be seen from Figure~\ref{fig: ph_check} that these lines are not ordered in such way and several lines cross each other. Hence, vine copula model (A) cannot be approximated by the Cox or AFT model.

Other examples in the next section show how copula-based models
can accommodate a wider variety of shapes of conditional quantile functions.

\section{Simulation Studies} \label{sec: simulation}


In this section, the effectiveness of the proposed prediction method for a censored response variable based on vine copulas is demonstrated on simulated datasets. We consider data generated from the AFT and vine copula models, apply these two methods to predict the censored response variable, and compare their performances. In time-to-event datasets collected from observational studies, there is often weak or moderate dependence among the explanatory variables. Since vine copulas are flexible models for multivariate dependence modeling, we assume the correlation among the explanatory variables can be captured by copula models when data are generated from the AFT model. The algorithm to simulate data from a vine copula model with continuous and discrete random variables has not been elaborated in any previous literature. The details of this algorithm are provided in the appendix. In all data generation scenarios, the sizes of the training and test sets are both 1000. The AFT and vine copula models are fitted on the training set. The evaluation metrics are computed based on the test set for each simulated dataset.

\subsection{Data Simulated from the AFT Model} \label{subsec: aft_simulation}

In this section, we consider scenarios where the time-to-event datasets are generated from the AFT model. Generalized Gamma distribution is fitted as the univariate distribution of $Y$ before fitting vine copulas. When computing the multiple imputed metrics, the estimated AFT model is used for conditional imputation.

\noindent\textit{Model (B). 5 dimensional AFT model with two discrete/ordinal explanatory variables}:


Explanatory variables 1 and 3 are continuous while explanatory variables 2 and 4 are discrete. The continuous univariate distributions are $N(0,1)$
and the discrete/ordinal variables are from discretized $N(0,1)$ random variables with categories $(-\infty,-1]$, $(-1,0]$, $(0,1]$, $(1,\infty)$. The response variable is generated according to $\log y_i = 0.08 x_{i1} + 0.06 x_{i2} + 0.09 x_{i3} + 0.07 x_{i4} + 0.2 W_i$. Correlations of the u-scores of the explanatory variables are specified by a 4 dimensional vine copula model. The vine array, bivariate copula family array, and bivariate copula parameter array of this vine copula model are in the appendix.

\noindent\textit{Model (C). 10 dimensional AFT model with four discrete/ordinal explanatory variables}:

Explanatory variables 1, 3, 5, 6, and 7 are continuous while explanatory variables 2 and 4 are discrete. The continuous univariate distributions are $N(0,1)$
and the discrete/ordinal variables are from discretized $N(0,1)$ random variables with cutpoints with categories $(-\infty,-1]$, $(-1,0]$, $(0,1]$, $(1,\infty)$. The response variable is generated according to $\log y_i = 0.04 x_{i1} + 0.05 x_{i2} + 0.04 x_{i3} + 0.03 x_{i4} + 0.04 x_{i5} + 0.05 x_{i6} + 0.03 x_{i7} + 0.1 W_i$. Correlations of the u-scores of the explanatory variables are specified by a 7 dimensional vine copula model. The vine array, bivariate copula family array, and bivariate copula parameter array of this vine copula model are in the appendix.

The MAE, RMSE and 50\% IS assuming that the true values of the censored responses are known as well as the multiple imputed MAE, RMSE and 50\% IS assuming that the true values of the censored responses are unknown at different censoring rates are summarized in Table~\ref{tab: sim_aft}. It can be seen the AFT model performs slightly better since data are simulated from the AFT model.

\begin{table}[!ht]
	\small
	\centering
	\begin{tabular}{*{8}{c}}
		\toprule
		\multicolumn{8}{c}{Model (B)}\\\midrule
		Censoring rate & Prediction method & MAE & RMSE & 50\% IS & MAE-MI & RMSE-MI & 50\% IS-MI\\\midrule
		\multirow{2}{*}{0.2} & AFT & 0.165 & 0.206 & 0.519 & 0.164 & 0.205 & 0.518 \\
		& Vine copula & 0.173 & 0.216 & 0.545 & 0.172 & 0.215 & 0.542 \\\midrule
		\multirow{2}{*}{0.5} & AFT & 0.165 & 0.206 & 0.520 & 0.165 & 0.206 & 0.520 \\
		& Vine copula & 0.173 & 0.218 & 0.549 & 0.172 & 0.216 & 0.545 \\\midrule
		\multirow{2}{*}{0.8} & AFT & 0.166 & 0.210 & 0.528 & 0.166 & 0.208 & 0.526 \\
		& Vine copula & 0.176 & 0.224 & 0.561 & 0.172 & 0.216 & 0.546 \\\midrule
		\multicolumn{8}{c}{Model (C)}\\\midrule
		Censoring rate & Prediction method & MAE & RMSE & 50\% IS & MAE-MI & RMSE-MI & 50\% IS-MI\\\midrule
		\multirow{2}{*}{0.2} & AFT & 0.097 & 0.124 & 0.308 & 0.096 & 0.123 & 0.305 \\
		& Vine copula & 0.101 & 0.125 & 0.320 & 0.100 & 0.124 & 0.317 \\\midrule
		\multirow{2}{*}{0.5} & AFT & 0.096 & 0.124 & 0.307 & 0.095 & 0.122 & 0.303 \\
		& Vine copula & 0.103 & 0.128 & 0.325 & 0.101 & 0.126 & 0.321 \\\midrule
		\multirow{2}{*}{0.8} & AFT & 0.096 & 0.124 & 0.320 & 0.095 & 0.122 & 0.319 \\
		& Vine copula & 0.103 & 0.128 & 0.333 & 0.101 & 0.126 & 0.332 \\
		\bottomrule
	\end{tabular}
	\caption{The absolute error (MAE), root mean squared error (RMSE), and interval score (IS) assuming that the true values of the censored responses are known as well as the multiple imputed MAE, RMSE and 50\% IS assuming that the true values of the censored responses are unknown comparing different methods in predicting the time-to-event response variable when at data are generated from the AFT models (B) and (C) at different censoring rates.}
	\label{tab: sim_aft}
\end{table}

\subsection{Data Simulated from the Vine Copula Model} \label{subsec: vine_simulation}


In this section, we consider scenarios where the time-to-event datasets are generated from vine copula models. 
Based on theory in \cite{chang2019prediction}, tail properties of
the linking copulas to the response affect the form of
conditional quantile functions at the extremes of the predictor space.
We consider many different scenarios of vine copulas to confirm that the
simulation performance matches what we expect from theory.
From this, we summarize results only for specific scenarios of linking copulas 
where we expect the AFT model to perform badly for predictions, as
this suffices to show that the vine copula approach has more flexibility.
When computing the multiple imputed metrics, the estimated vine copula model is used for conditional imputation.

\noindent\textit{Model (D). 5 dimensional vine copula model with two discrete/ordinal explanatory variables}:

Explanatory variables 2 and 3 are continuous while explanatory variables 1 and 4 are discrete. The univariate distributions are $\text{Exp}(1)$ for the continuous explanatory variables and the discrete/ordinal variables are from discretized $\text{Exp}(1)$ random variables with categories $(0,0.35]$, $(0.35,0.7]$, $(0.7,1.25]$, $(1.25,\infty)$. The discrete values are set to 0.2, 0.5, 0.9, and 1.6 for the four categories. The response variable follows log-normal distribution. Correlations of the u-scores of the explanatory variables and the response variable are specified by a 5 dimensional vine copula model. The vine array, bivariate copula family array, and bivariate copula parameter array of this vine copula model are in the appendix.

\noindent\textit{Model (E). 8 dimensional vine copula model with four discrete/ordinal explanatory variables}:

Explanatory variables 1, 2, 3, 4, and 5 are continuous while explanatory variables 6 and 7 are discrete. The univariate distributions are $\text{Exp}(1)$ for the continuous explanatory variables and the discrete/ordinal variables are from discretized $\text{Exp}(1)$ random variables with categories $(0,0.35]$, $(0.35,0.7]$, $(0.7,1.25]$, $(1.25,\infty)$. The discrete values are set to 0.2, 0.5, 0.9, and 1.6 for the four categories. The response variable follows log-normal distribution. Correlations of the u-scores of the explanatory variables and the response variable are specified by an 8 dimensional vine copula model. The vine array, bivariate copula family array, and bivariate copula parameter array of this vine copula model are in the appendix.

In these two simulation cases, the bivariate copula models associated with the censored response are all set to the Frank copula which has tail quadrant
independence. The use of such linking copulas implies that conditional  
quantile functions are flattening at the extremes of the predictor space.
When some of linking copulas have stronger dependence in the joint tails, the conditional quantile functions can be increasing at different rates
as one explanatory variable increases with others held fixed.

We apply our proposed algorithms in Section~\ref{sec: survival_copula} to estimate the vine copula structure and make predictions for the response based on the fitted vine copula model. The MAE, RMSE and 50\% IS assuming that the true values of the censored responses are known as well as the multiple imputed MAE, RMSE and 50\% IS assuming that the true values of the censored responses are unknown at different censoring rates are summarized in Table~\ref{tab: sim_vine}. It can be seen that the vine copula model performs much better than the AFT model at all censoring rates. In these two cases, the univariate distributions for the explanatory variables and the response variable are exponential and log-normal. Therefore, the explanatory variables and the response variable are skewed to the right and may take extreme values. It can also be easily verified that the proportional hazards or AFT assumptions do not hold in these two cases. Table~\ref{tab: sim_vine} shows that as the censoring rate increases, the MAE, RMSE, and 50\% IS of the AFT model become larger and larger, indicating that the AFT model fails to fit the data well with extreme observations at high censoring rates. As the censoring rate increases, the absolute values of the estimates of the intercept and slopes of the AFT model become larger, leading to over-predicted medians and quantiles across all data points.

\begin{table}[!ht]
	\small
	\centering
	\begin{tabular}{*{8}{c}}
		\toprule
		\multicolumn{8}{c}{Model (D)}\\\midrule
		Censoring rate & Prediction method & MAE & RMSE & 50\% IS & MAE-MI & RMSE-MI & 50\% IS-MI\\\midrule
		\multirow{2}{*}{0.2} & AFT & 1.302 & 3.284 & 4.209 & 1.265 & 3.243 & 4.118 \\
		& Vine copula & 0.892 & 1.770 & 2.915 & 0.826 & 1.957 & 2.800 \\\midrule
		\multirow{2}{*}{0.5} & AFT & 2.368 & 8.624 & 7.545 & 2.363 & 8.502 & 7.462 \\
		& Vine copula & 0.865 & 1.957 & 2.781 & 0.723 & 1.669 & 2.573 \\\midrule
		\multirow{2}{*}{0.8} & AFT & 3.747 & 13.69 & 10.01 & 3.834 & 13.73 & 10.11 \\
		& Vine copula & 0.729 & 1.372 & 2.365 & 0.820 & 1.516 & 2.674 \\\midrule
		\multicolumn{8}{c}{Model (E)}\\\midrule
		Censoring rate & Prediction method & MAE & RMSE & 50\% IS & MAE-MI & RMSE-MI & 50\% IS-MI\\\midrule
		\multirow{2}{*}{0.2} & AFT & 0.994 & 1.771 & 3.312 & 1.025 & 2.064 & 3.417 \\
		& Vine copula & 0.892 & 1.770 & 2.972 & 0.899 & 2.089 & 3.012 \\\midrule
		\multirow{2}{*}{0.5} & AFT & 1.295 & 3.048 & 4.161 & 1.348 & 3.156 & 4.310 \\
		& Vine copula & 0.897 & 1.785 & 2.978 & 0.893 & 2.064 & 3.015 \\\midrule
		\multirow{2}{*}{0.8} & AFT & 19.52 & 286.4 & 65.09 & 19.55 & 286.3 & 65.24 \\
		& Vine copula &  0.908 & 1.802 & 3.012 & 0.712 & 1.683 & 2.402 \\
		\bottomrule
	\end{tabular}
	\caption{The absolute error (MAE), root mean squared error (RMSE), and interval score (IS) assuming that the true values of the censored responses are known as well as the multiple imputed MAE, RMSE and 50\% IS assuming that the true values of the censored responses are unknown comparing different methods in predicting the time-to-event response variable when at data are generated from the vine copula models (D) and (E) at different censoring rates.}
	\label{tab: sim_vine}
\end{table}

\begin{figure}[ht] 
	\centering
	\begin{subfigure}[b]{0.32\textwidth}
		\centering
		\includegraphics[width=\textwidth]{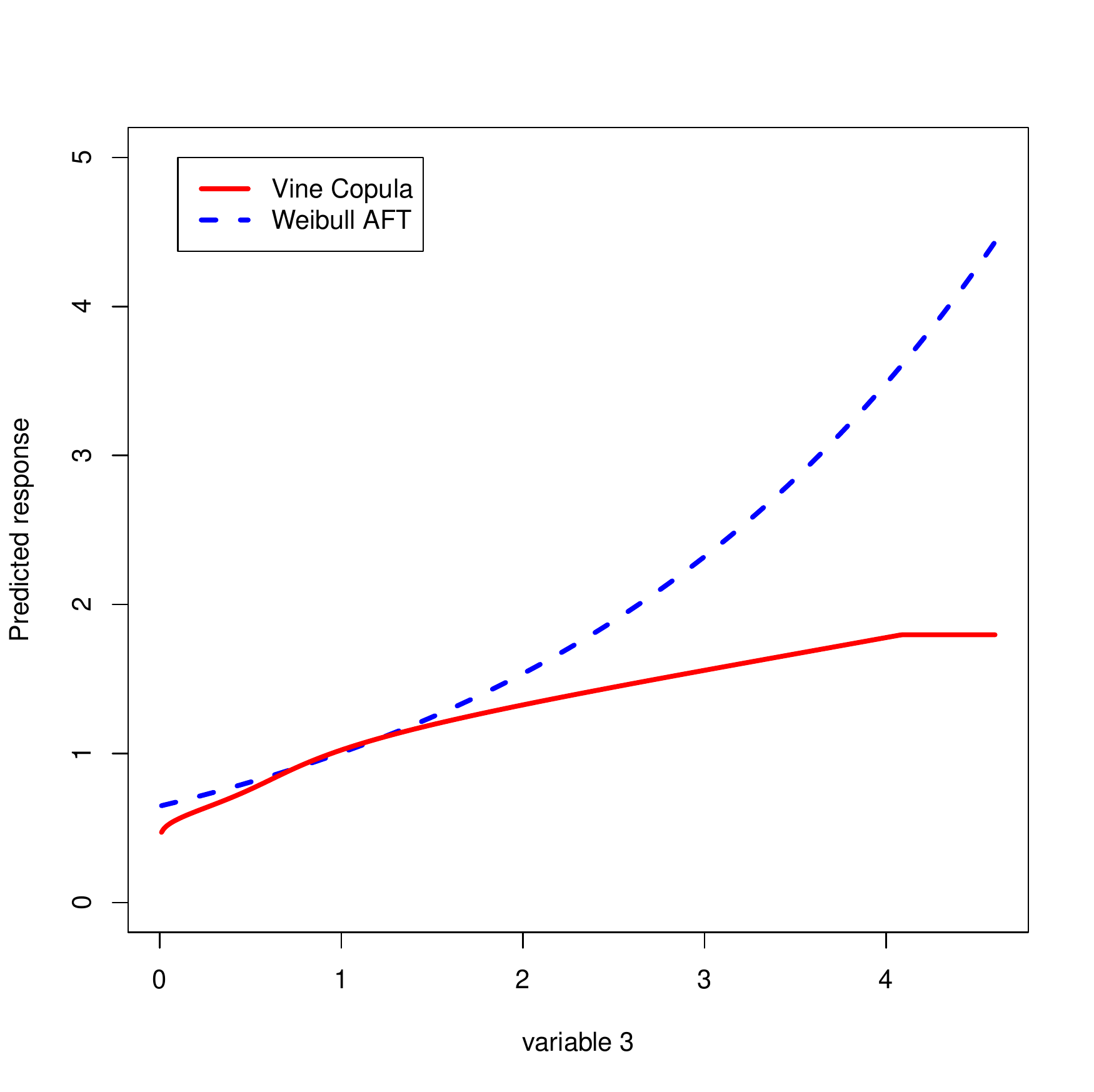}
	\end{subfigure}
	\hfill
	\begin{subfigure}[b]{0.32\textwidth}
		\centering
		\includegraphics[width=\textwidth]{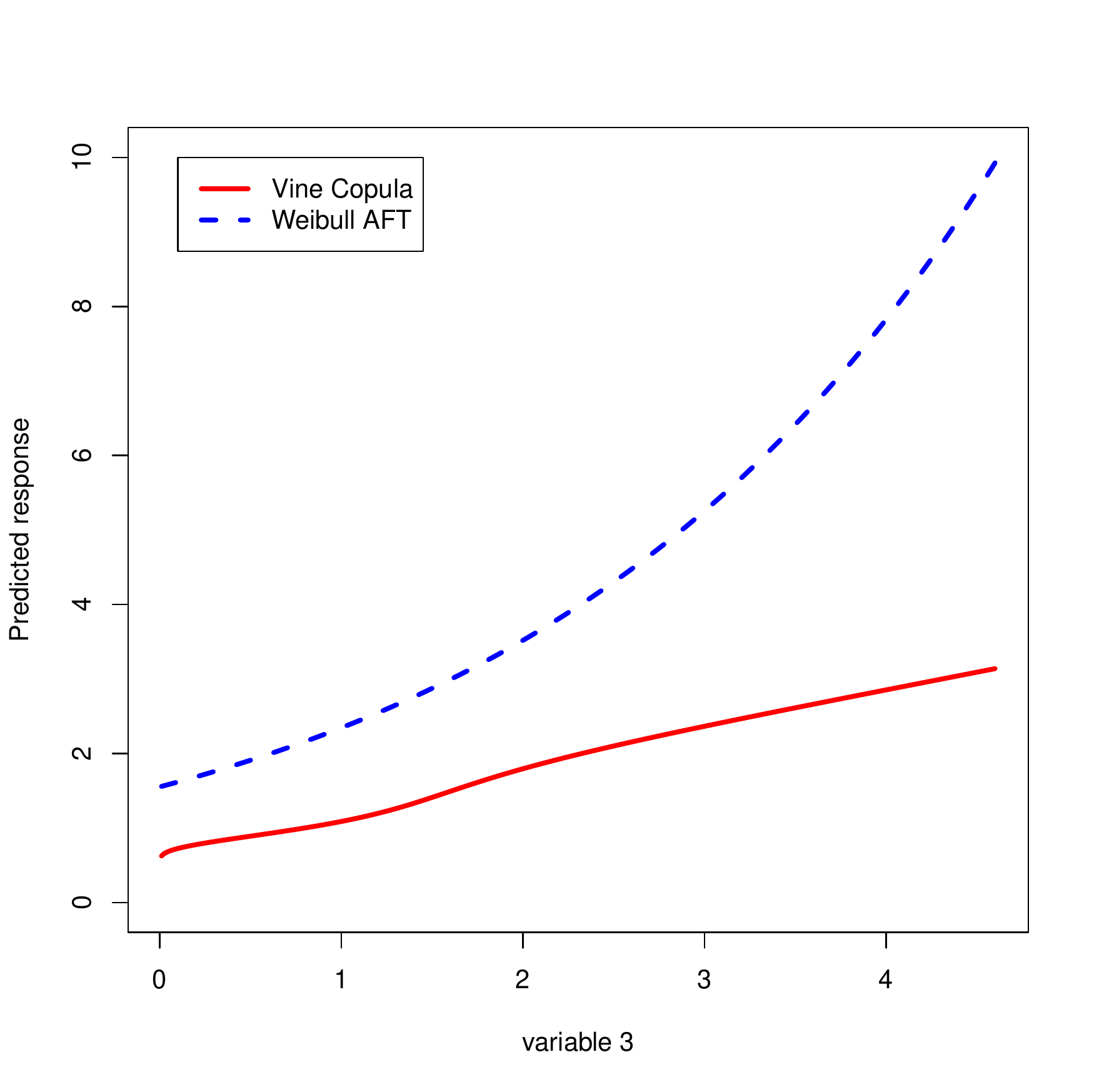}
	\end{subfigure}
	\hfill
	\begin{subfigure}[b]{0.32\textwidth}
		\centering
		\includegraphics[width=\textwidth]{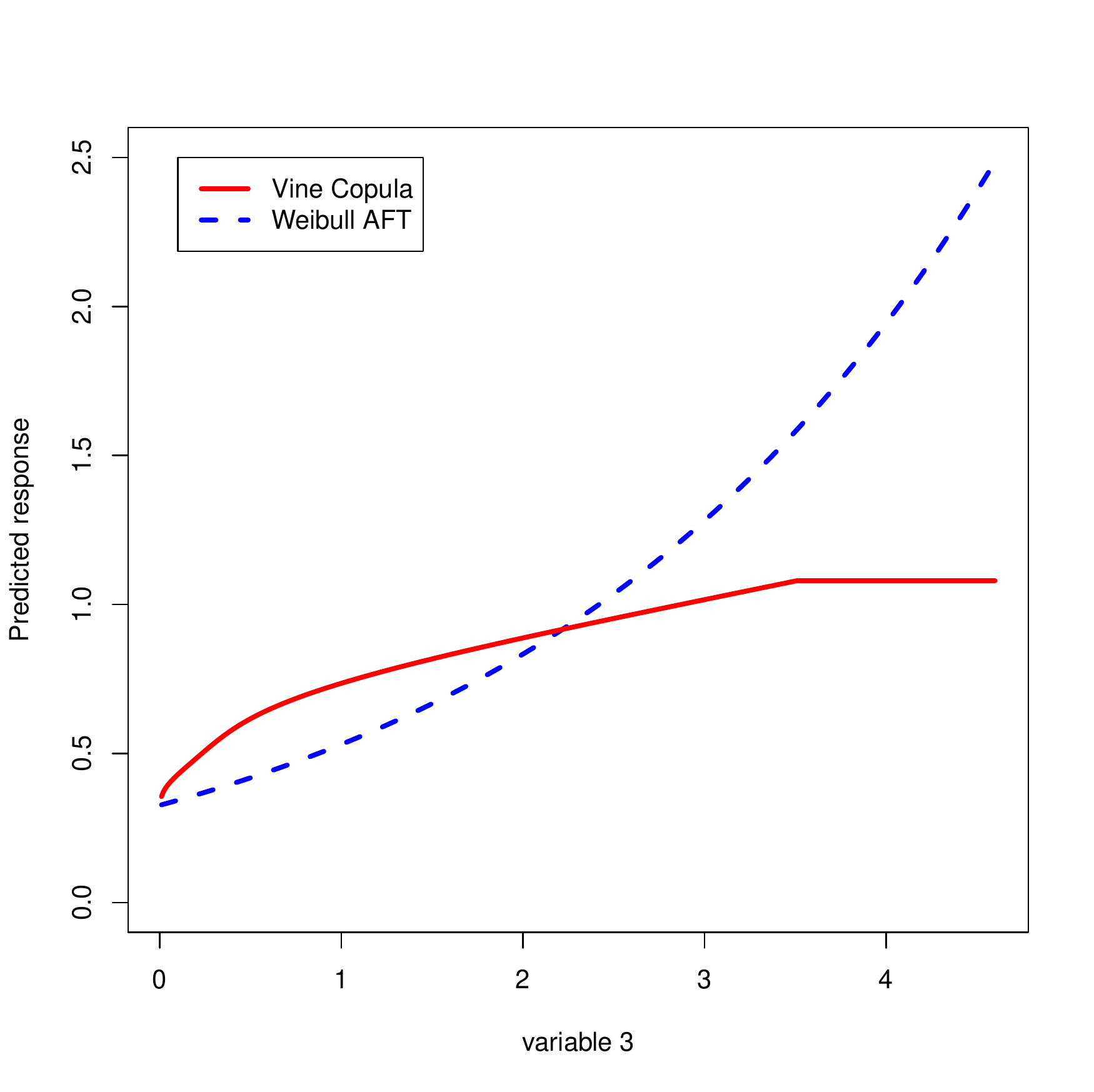}
	\end{subfigure}
	\caption{Plots of the median response predicted by the AFT and vine copula models with varying variable 3 from 0.01 to 0.99 quantiles for three randomly selected observations in the training set for Model (D) with censoring rate 0.2. All the other predictors are set to those of the selected observations.}
	\label{fig: sim}
\end{figure}

To further compare the predictions given by the AFT and vine copula models, we change one explanatory variable and fix the rest and examine how predictions of the two models vary. We take Model (D) with censoring rate 0.2 as an example. We make variable 3 a varying predictor and set the values of the other predictors to one randomly chosen observation in the training set. The median responses predicted by the AFT and vine copula models with varying variable 3 from 0.01 to 0.99 quantiles for three randomly selected observations are plotted in Figure~\ref{fig: sim}. The true values for the uncensored response variable are 0.375, 0.841, and 0.724, respectively. Even with a varying predictor, the predicted medians should not deviate too much from the true values. However, from Figure~\ref{fig: sim}, it can be seen that the AFT model tends to make extreme predictions given large predictors while predictions from the vine copula model tend to flatten out as a single predictor increases. This shows that the vine copula approach is more flexible when the AFT assumptions are violated as the response does not have a log-linear relationship with the explanatory variables.
Also note that there is no general approach in the proportional
hazards model to replace
$\x^T\gbf$ by $g(\x,\thbf)$ to have more non-linear shapes as the predictor
space expands beyond a local region.
In other simulation scenarios (not shown here), the prediction performance
of AFT relative to copula-based is a little worse
when the linking copulas to the response have stronger dependence in the
tails, leading to conditional quantile functions that continue to
increase for large values of explanatory variables.

Im summary, the vine copula modeling approach can provide a decent approximation to other time-to-event models. When data are generated from the AFT model, the vine copula method works reasonably well in predicting the censored response. In contrast, many conditional quantile functions from vine copula models cannot be well approximated by the Cox or AFT models with a linear function of explanatory variables. When data are generated from vine copula models, the vine copula method can perform much better (depending on the shape of conditional quantiles) than the AFT model in predicting the censored response variable. Therefore, our proposed method of predicting event times with vine copulas is a practical alternative to traditional time-to-event models when the predictor space can be considered as unbounded.

\section{Application to a Survival Dataset} \label{sec: app}

In this section, we compare the prediction performance of our proposed vine copula approach with traditional survival analysis models on the primary biliary cirrhosis (PBC) dataset introduced in \cite{fleming2011counting}. This dataset is available in the \verb|survival| R package (\cite{therneau2015package}).

Primary sclerosing cholangitis is an autoimmune disease leading to destruction of the small bile ducts in the liver. Progression is slow but inexhortable, eventually leading to cirrhosis and liver decompensation. This dataset was collected from the Mayo Clinic trial conducted between 1974 and 1984. There are 312 patients participated in the randomized trial. We select seven explanatory variables, age, edema, bili (serum bilirunbin), albumin (serum albumin), platelet (platelet count), protime (standardized blood clotting time), and stage (histologic stage of disease), which are largely complete, as the explanatory variables to predict the survival times (in days) of the participants. Among all explanatory variables, edema and stage are discrete; all the others are continuous. We replace the four missing explanatory variables in the dataset by the sum of the median value and a small Gaussian random noise. The status of each participant at the endpoint of the trial can fall into one of the three categories: censored, transplant, or dead. In this analysis, we treat participants from the first two categories as right-censored and participants from the last category as fully observed. We randomly select 240 cases as the training set and the remaining 72 cases are used as the test set. We tried multiple random splits of the training/test sets. The results of different splits are similar to those reported in the remainder of this section.

The first step of applying the vine copula approach is to fit the univariate distributions for each variable and then transform them to u-scores. For the continuous explanatory variables, we fit the normal distribution for age, the Weibull distribution for bili, platelet, and protime, and the skewed normal distribution for albumin based on the training data. Diagnostic plots suggest that these models fit the variables adequately. For the discrete explanatory variables, we apply the rank transform based on the training set to obtain the u-scores. For the response variable, we fit a left-truncated normal distribution with minimum value 0 and optimize the parameters based on the log-likelihood of both censored and uncensored response in the training set.
The Q-Q plots show that the left-truncated normal distribution fits the response variable better than other candidates. The parameter estimates of the univariate distributions for the continuous explanatory variables and the response variable as well as the estimated van der Waerden correlations with the response are displayed in Table~\ref{tab: univariate}. For the u-scores of age, edema, bili, protime, and stage, we further take a $u' = 1-u$ transform to make sure that they have positive van der Waerden correlations with the response variable. Details of the fitted vine array and bivariate copula family array can be found in the appendix.

\begin{table}[!ht]
\small
\centering
\begin{tabular}{lllc}
\toprule
Variable & Distribution & Parameter Estimates & vW cor \\\midrule
age & Normal & $\mu$: 49.9, $\sigma$: 10.5 & -0.34 \\
bili & Weibull & $k$ (shape): 0.895, $\lambda$ (scale): 3.04 & -0.63 \\
albumin & Skewed normal & $\xi$ (location): 3.51, $\omega$ (scale): 0.426, $\alpha$ (shape): 0.00702 & 0.45 \\
platelet & Weibull & $k$ (shape): 3.01, $\lambda$ (scale): 295 & 0.24 \\
protime & Weibull & $k$ (shape): 9.71, $\lambda$ (scale): 11.2 & -0.52 \\
time (response) & Truncated normal & $\mu$: 554, $\sigma$: 4609, $a$ (lower limit): 0, $b$ (upper limit): $\infty$ & N/A \\		
\bottomrule
\end{tabular}
\caption{The fitted univariate distributions and the corresponding parameter estimates for the continuous explanatory variables and the response variable as well as the estimated van der Waerden correlations with the response based on the training set.}
\label{tab: univariate}
\end{table}

For the AFT model, we take the log-transformation of two explanatory variables bili and protime to make them less skewed. We then rescale all the explanatory variables and the response variable using min-max normalization to normalize their ranges. The coefficient estimates from the AFT model for the seven min-max and/or log transformed explanatory variables are -0.144 for age, -0.082 for edema, -0.431 for bili, 0.117 for albumin, 0.004 for platelet, -0.087 for protime, and -0.110 for stage, respectively, indicating that most predictors are negatively correlated with the survival time.

For comparison purposes, we also consider Cox partial likelihood
estimation with Weibull baseline hazard function, as mentioned in
Section \ref{subsec: cox}.
This results in similar estimates to maximum likelihood with the AFT
model.
We take the same preprocessing steps as the AFT model. The coefficient estimates from the Cox model for the seven min-max and/or log transformed explanatory variables are 0.357 for age, 0.207 for edema, 0.993 for bili, -0.294 for albumin, -0.010 for platelet, 0.230 for protime, and 0.247 for stage, respectively. Note that the estimates from the Cox model have different signs but are approximately proportional to those from the AFT model.

To compare the prediction performances on the censored data, a conditional distribution $F_{Y|\X}$ is needed to generate multiple sets of imputed values to evaluate our proposed metrics in Section~\ref{subsubsec: metrics_survival}. One of the features of the PBC dataset is that many censored values of the response variable are close to or greater than the largest observed value. This makes it unreliable to make predictions for these large censored values since the upper tail is completely unobserved. Under this circumstance, a model with light upper tail for the response variable should be preferred to make more conservative predictions for the censored cases. If we use $F_{Y|\X}$ derived from the fitted AFT or Cox model as the conditional distribution, the imputed values tend to have very heavy upper tails, leading to the prediction intervals generated by all the three models having exceptionally low coverage rates. In contrast, the prediction intervals generated by the three methods have desired coverage rates when they are evaluated on the imputed values simulated from the fitted vine copula model; the upper tails of the fitted conditional distribution are also lighter. Therefore, we conclude that the vine copula model is a more suitable model for the PBC dataset and is thus used for multiple imputation. The MAE, RMSE and 50\% IS for the uncensored data in the test set as well as the multiple imputed MAE, RMSE and 50\% IS combining the uncensored and imputed data in the test set for the AFT, Cox, and vine copula models using all the explanatory variables are summarized in Table~\ref{tab: pbc}. It can be seen the vine copula prediction approach has superior performances compared with the AFT and Cox models on both uncensored and imputed data. This shows the effectiveness of our proposed vine copula survival prediction method. Since the prediction results generated by the Cox model are very similar to the AFT model and they have similar parametric forms, we only compare the vine copula approach with the AFT model in the remainder of this section.

\begin{table}[!ht]
	\small
	\centering
	\begin{tabular}{*{7}{c}}
		\toprule
		Prediction method & MAE & RMSE & 50\% IS & MAE-MI & RMSE-MI & 50\% IS-MI\\\midrule
		AFT & 1158 & 1694 & 3539 & 1500 & 2057 & 4851\\
		Cox & 1136 & 1650 & 3470 & 1477 & 2034 & 4810 \\
		Vine copula & 994 & 1495 & 3311 & 1407 & 1990 & 4623 \\
		\bottomrule
	\end{tabular}
	\caption{The absolute error (MAE), root mean squared error (RMSE), and interval score (IS) for the uncensored data only in the test set as well as the multiple imputed MAE, RMSE and 50\% IS combining the uncensored and imputed data in the test set for the AFT, Cox, and vine copula models using all the explanatory variables.}
	\label{tab: pbc}
\end{table}

\begin{figure}[ht] 
	\centering
	\begin{subfigure}[b]{0.32\textwidth}
		\centering
		\includegraphics[width=\textwidth]{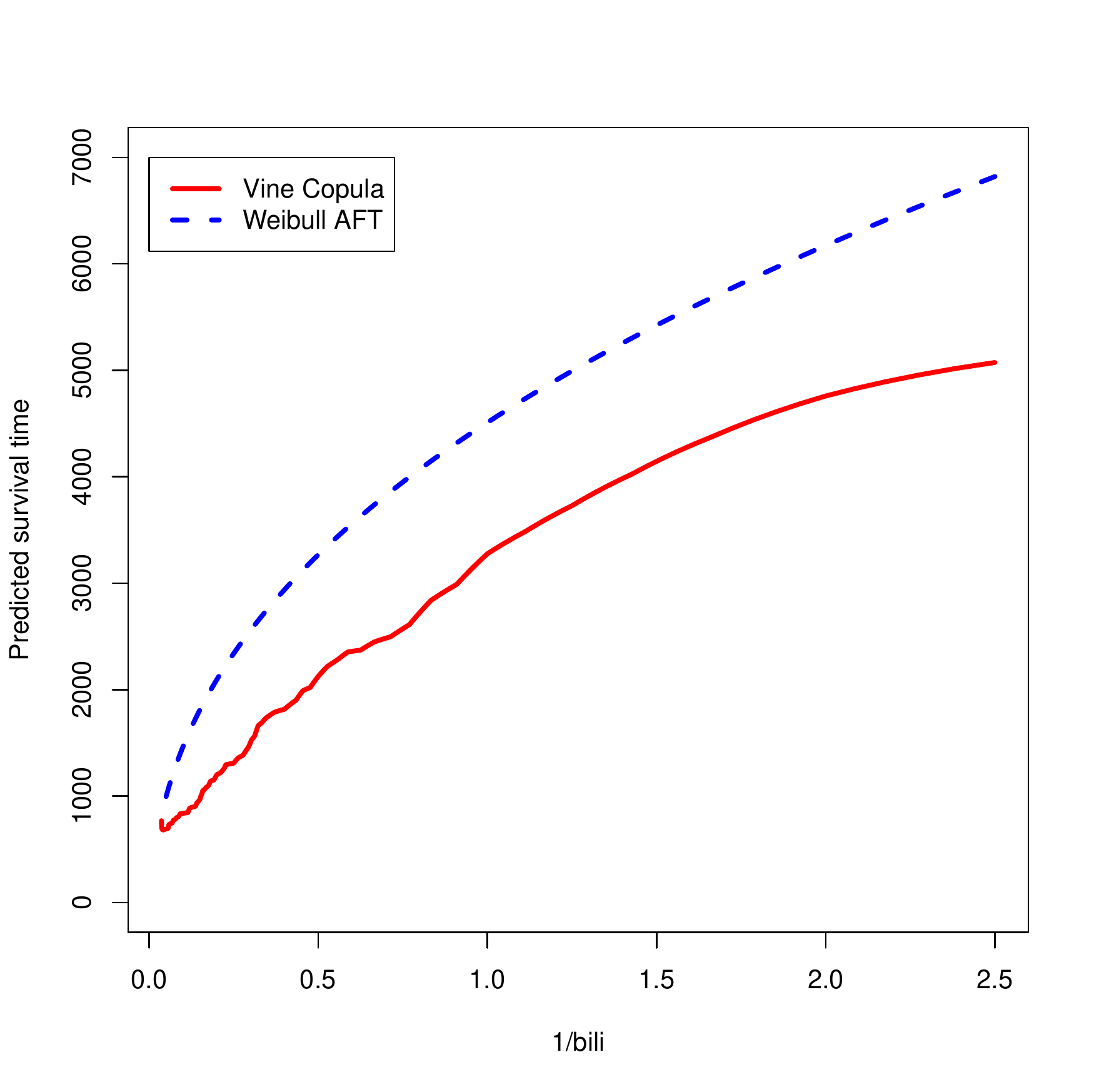}
	\end{subfigure}
	\hfill
	\begin{subfigure}[b]{0.32\textwidth}
		\centering
		\includegraphics[width=\textwidth]{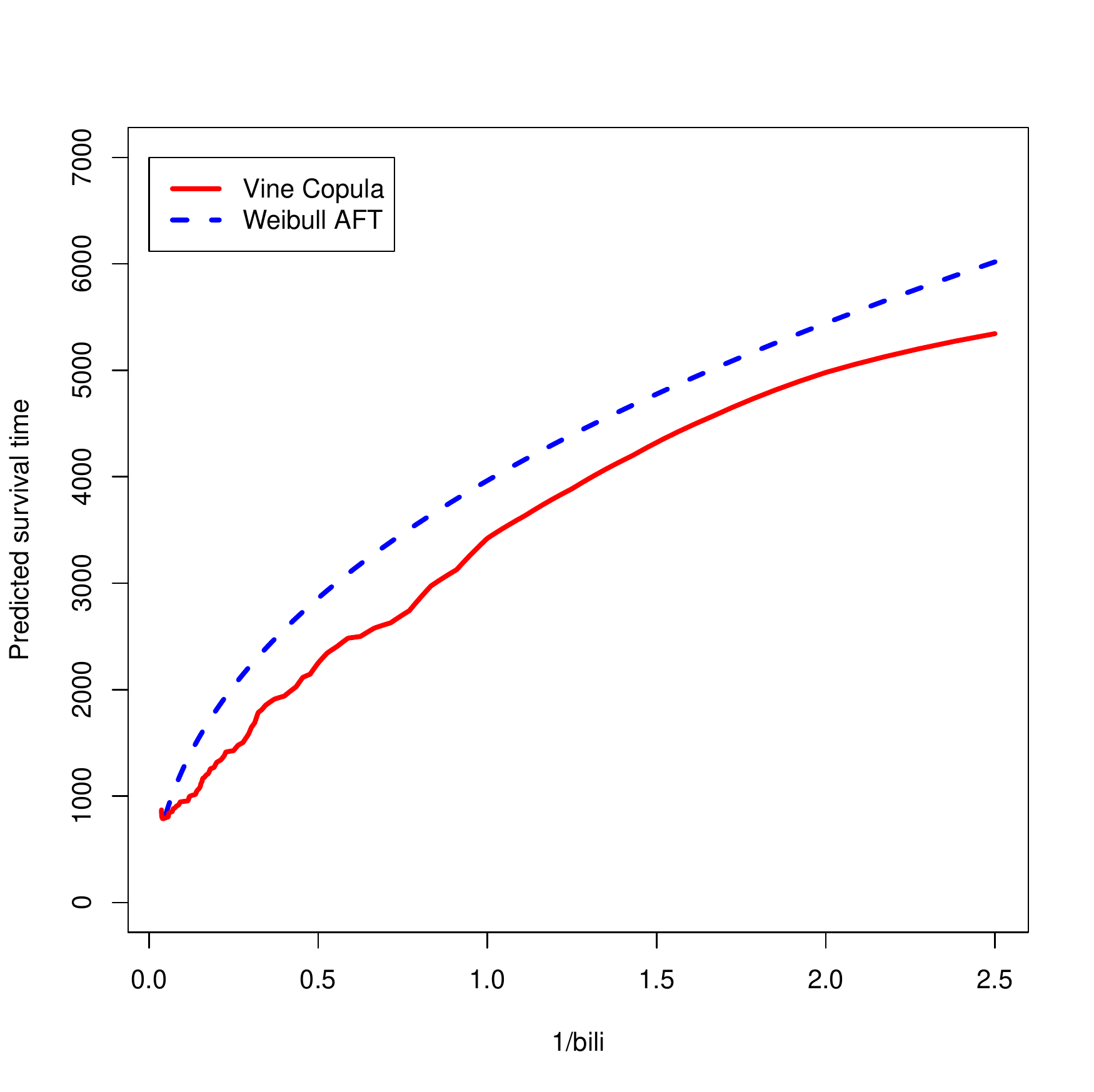}
	\end{subfigure}
	\hfill
	\begin{subfigure}[b]{0.32\textwidth}
		\centering
		\includegraphics[width=\textwidth]{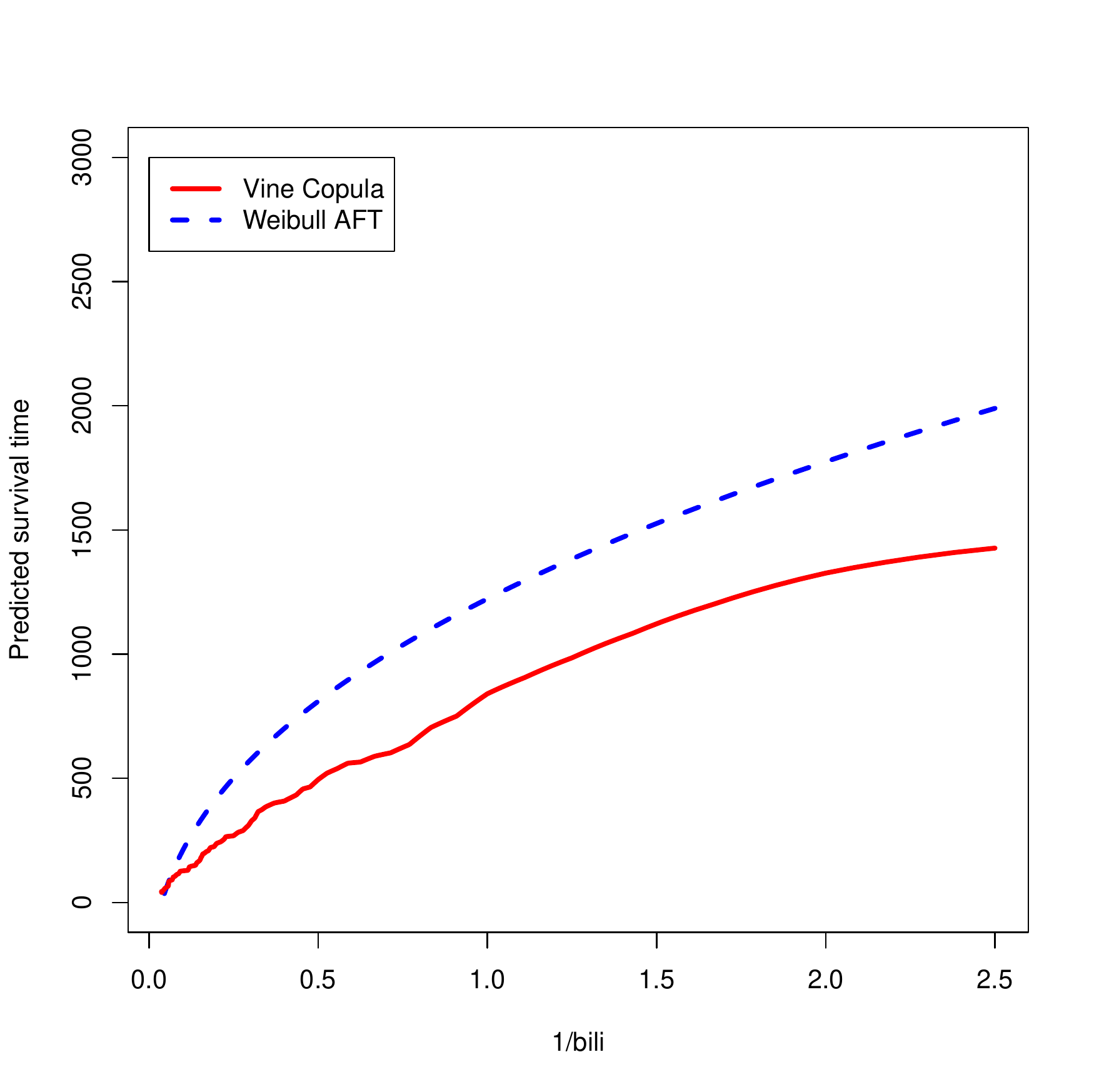}
	\end{subfigure}
	\caption{Plots of the median survival times predicted by the AFT and vine copula models with varying bili from 0.005 to 0.995 quantiles for three randomly selected observations in the training set. All the other predictors are set to those of the selected observations.}
	\label{fig: bili}
\end{figure}

To further compare the predictions given by the AFT and vine copula models, we change one explanatory variable and fix the rest and examine how predictions of the two models vary. We take the continuous variable bili as an example of varying predictor and set the values of the other predictors to one randomly chosen observation in the training set. The median survival times predicted by the AFT and vine copula models with varying bili from 0.005 to 0.995 quantiles (plotted as 1/bili on the x-axis since it is negatively correlated with the response) for three randomly selected observations are plotted in Figure~\ref{fig: bili}. It can be seen that when bili takes more extreme values (i.e., small values close to 0), the vine copula model tends to generate more conservative predictions since bili is linked to the response variable by a Frank copula. Since the upper tail behaviors of the survival time are only partially observed in the dataset, conservative predictions should be preferred even with extreme predictors.

Moreover, as a sensitivity analysis, we also compare the prediction performances of the two methods when different sets of explanatory variables are used. The results are summarized in Table~\ref{tab: sensitivity}. For the case of no predictors, the baseline method always predicts the median of the fitted truncated normal distribution. For the one explanatory variable case, bili, which has the largest absolute van der Waerden correlation with the response, is used as the single predictor. Bivariate Frank copula is used as the copula model for this case since it fits the data better than other bivariate copula families. For the five explanatory variable case, all the continuous variables, i.e., age, bili, albumin, platelet, and protime, are used as the predictors. It can be seen that in both one and five variable cases, the AFT and vine copula methods have significantly superior prediction performances compared with the baseline method in terms of point and interval predictions. Moreover, the copula model consistently outperforms the AFT model, showing the flexibility of the copula approach.

\begin{table}[!ht]
	\small
	\centering
	\begin{tabular}{*{8}{c}}
		\toprule
		Variable & Prediction method & MAE & RMSE & 50\% IS & MAE-MI & RMSE-MI & 50\% IS-MI\\\midrule
		None & Median & 1980 & 2209 & 6278 & 2290 & 2912 & 7170 \\\midrule
		\multirow{2}{*}{1 (bili)} & AFT & 1377 & 1759 & 4204 & 1681 & 2260 & 5490 \\
		& Bivariate Frank copula & 1275 & 1591 & 3922 & 1668 & 2256 & 5336 \\\midrule
		\multirow{2}{*}{5 (all continuous)} & AFT & 1215 & 1687 & 3678 & 1507 & 2033 & 4892 \\
		& Vine copula & 1097 & 1550 & 3537 & 1494 & 2069 & 4782 \\
		\bottomrule
	\end{tabular}
	\caption{The absolute error (MAE), root mean squared error (RMSE), and interval score (IS) for the uncensored data only in the test set as well as the multiple imputed MAE, RMSE and 50\% IS combining the uncensored and imputed data in the test set for the AFT and vine copula models using different sets of explanatory variables.}
	\label{tab: sensitivity}
\end{table}

\section{Conclusion} \label{sec: conclude}


In conclusion, we propose a vine copula based approach to predict a right-censored response variable given a set of mixed discrete-continuous explanatory variables. Compared with other traditional time-to-event analysis methods such as Cox proportional hazards or AFT models, our proposed method provides
conditional quantile functions that lead to more accurate point predictions and more informative interval predictions, and can often approximate other models well. In contrast, conditional quantile functions of vine copulas usually cannot be approximated by other time-to-event analysis models. When the proportional hazards or AFT assumptions are not satisfied, predictions based on vine copulas can significantly outperform other models depending on the shapes of conditional quantile functions. We also adapt common metrics to evaluate the prediction performances on censored datasets by applying multiple imputation. Simulation studies demonstrate the effectiveness of our proposed prediction methods and evaluation metrics. Finally, we apply our proposed method to the PBC dataset and achieve superior prediction performances compared with other methods.

In this paper, the proposed vine copula prediction method specifically works for right-censored data. With minor modifications to the likelihood function, our proposed method can be extended to predictions of other types of censored response, such as left and interval-censored data.

\section*{Acknowledgments}
This research has been supported by NSERC Discovery Grant 8698. 

\bibliographystyle{apalike}
\bibliography{literature}

\pagebreak
\section*{Appendix}
\appendix
\section{Algorithm for Simulating Data from Vine Copulas with Continuous and Discrete Variables}
In this section, we state the algorithm for simulating data from a vine copula model with continuous and discrete random variables in Algorithm~\ref{alg: simulation}. The inputs to the algorithm are the vine array $A_{d\times d} = (a_{kj})$ with $a_{jj} = j$ for $j = 1, \dots, d$ on the diagonal, the bivariate copula family matrix $F_{d\times d}$ for all the edges in the vine. For variable $j$, if it is discrete, we assume it is discretized by cutpoints $\infty, \text{cutPoints}_1, \text{cutPoints}_2, \dots, \text{cutPoints}_{m-1}, \infty$ with categories $(-\infty, \text{cutPoints}_1]$, $(\text{cutPoints}_1, \text{cutPoints}_2]$, $\dots$, $(\text{cutPoints}_{m-1}, \infty]$. In each of these categories, we assume the discrete variable takes value $\text{discVal}_{j,1}, \text{discVal}_{j,2}, \dots,$, and $\text{discVal}_{j,m}$, respectively. The discrete values can be set to the middle points of these categories. The output of the algorithm is an $n$ by $d$ matrix of the simulated values from the specified vine copula model.

\begin{algorithm}
	\caption{Simulating $n$ sets of observations from an R-vine with continuous or discrete explanatory variables as variables $1, \dots, d$. Inputs are the univariate CDFs $F_{X_j}$ for $j = 1, \dots, d$ as well as the cutpoints $\text{cutPoints}_{j, 1:(m+1)}$ and discrete values $\text{discVal}_{j,1:m}$ for variable $j$ if it is discrete.
	}
	\footnotesize
	\begin{algorithmic}[1]
		\State Initialize $X = (x_{ij})$ as an empty matrix, $i = 1,\dots, n$, $j = 1, \dots, d$.
		\State Simulate $U = (u_{ij})$ where $u_{ij}~\overset{\text{i.i.d.}}{\sim}U(0,1)$ for $i = 1,\dots, n$, $j = 1, \dots, d$. 
		\If{isDiscrete(variable $a_{11}$)}
		\State Compute $F_{X_{a_{11}}}(\text{cutPoints}_{a_{11},1}), \dots, F_{X_{a_{11}}}(\text{cutPoints}_{a_{11},m+1})$.
		\For{$i = 1,\dots,n$}
		\State Set $x_{i,a_{11}} \gets \text{discVal}_{a_{11},k}$ if $F_{X_{a_{11}}}(\text{cutPoints}_{a_{11},k}) \leq u_{i,a_{11}} < F_{X_{a_{11}}}(\text{cutPoints}_{a_{11},k+1})$.
		\EndFor
		\Else
		\For{$i = 1,\dots,n$}
		\State Set $x_{i,a_{11}} \gets F_{X_{a_{11}}}^{-1}\left(u_{i,a_{11}}\right)$.
		\EndFor
		\EndIf
		\For{$j = 2,\dots,d$}
		\If{isDiscrete(variable $a_{jj}$)}
		\For{$i = 1,\dots,n$}
		\State Compute $F_{X_{a_{jj}}|X_{a_{11},\dots,a_{j-1,j-1}}}\left(\text{cutPoints}_{a_{jj},k}|x_{i,a_{11}},\dots,x_{i,a_{j-1,j-1}}\right)$ for $k = 1, \dots, m+1$.
		\State Set $x_{i,a_{jj}} \gets \text{discVal}_{a_{jj},k}$ if $F_{X_{a_{jj}}|X_{a_{11},\dots,a_{j-1,j-1}}}\left(\text{cutPoints}_{a_{jj},k}|x_{i,a_{11}},\dots,x_{i,a_{j-1,j-1}}\right) \leq u_{i,a_{11}} < F_{X_{a_{jj}}|X_{a_{11},\dots,a_{j-1,j-1}}}\left(\text{cutPoints}_{a_{jj},k+1}|x_{i,a_{11}},\dots,x_{i,a_{j-1,j-1}}\right)$.
		\EndFor
		\Else
		\For{$i = 1,\dots,n$}
		\State Set $x_{i,a_{jj}} \gets F_{X_{a_{jj}}|a_{11},\dots,a_{j-1,j-1}}^{-1}\left(u_{i,a_{jj}}|x_{i,a_{11}, \dots, x_{i, a_{j-1,j-1}}}\right)$ by solving $F_{X_{a_{jj}}|a_{11},\dots,a_{j-1,j-1}}\left(\cdot|x_{i,a_{11}, \dots, x_{i, a_{j-1,j-1}}}\right) = u_{i,a_{jj}}$ using Algorithm 3.1 in \cite{chang2019prediction}.
		\EndFor
		\EndIf
		\EndFor
		\State Return $X$.
	\end{algorithmic} \label{alg: simulation}
\end{algorithm}

\section{Details of the Simulation Models}
The vine array, bivariate copula family array, and bivariate copula parameter array for model (B) in Section~\ref{subsec: aft_simulation} are as follows:
\[
A = 
\begin{bmatrix}
1 & 1 & 1 & 1\\
& 2 & 2 & 2\\
&   & 3 & 3\\
&   &   & 4\\
\end{bmatrix},
F =
\begin{bmatrix}
- & \text{t} & \text{t} & \text{t}\\
& - & \text{N} & \text{N}\\
& & - & \text{N}\\
& & & - \\
\end{bmatrix},
P =
\begin{bmatrix}
- & 0.49 (6) & 0.46 (6) & 0.50 (6)\\
& - & 0.22 & 0.34\\
& & - & 0.05\\
& & & - \\
\end{bmatrix},
\]
where N, t, F, and G in the copula family matrix stands for bivariate Gaussian (normal), t, Frank, and Gumbel copulas, respectively.

The vine array, bivariate copula family array, and bivariate copula parameter array for model (C) in Section~\ref{subsec: aft_simulation} are as follows:
\[
A = 
\begin{bmatrix}
1 & 1 & 1 & 1 & 1 & 1 & 1\\
& 2 & 2 & 2 & 2 & 2 & 2\\
&   & 3 & 3 & 3 & 3 & 3\\
&   &   & 4 & 4 & 4 & 4\\
&   &   &   & 5 & 5 & 5\\
&   &   &   &   & 6 & 6\\
&   &   &   &   &   & 7\\
\end{bmatrix},
F =
\begin{bmatrix}
- & \text{t} & \text{t} & \text{t} & \text{t} & \text{t} & \text{t} \\
& - & \text{t} & \text{t} & \text{t} & \text{t} & \text{t} \\
&   & - & \text{N} & \text{N} & \text{N} & \text{N} \\
&   &   & - & \text{N} & \text{N} & \text{N} \\
&   &   &   & - & \text{N} & \text{N} \\
&   &   &   &   & - & \text{N} \\
&   &   &   &   &   & - \\
\end{bmatrix},
\]
\[
P =
\begin{bmatrix}
- & 0.64 (6) & 0.60 (6) & 0.65 (6) & 0.57 (6) & 0.65 (6) & 0.54 (6) \\
& - & 0.40 (8) & 0.45 (8) & 0.45 (8) & 0.38 (8) & 0.30 (8) \\
&   & - & 0.12 & 0.25 & 0.23 & 0.29 \\
&   &   & - & 0.01 & 0.07 & 0.09 \\
&   &   &   & - & 0.03 & 0.09 \\
&   &   &   &   & - & 0.03 \\
&   &   &   &   &   & - \\
\end{bmatrix}.
\]

The vine array, bivariate copula family array, and bivariate copula parameter array for model (D) in Section~\ref{subsec: vine_simulation} are as follows:
\[
A = 
\begin{bmatrix}
1 & 1 & 3 & 3 & 4\\
& 3 & 1 & 2 & 3\\
&   & 2 & 1 & 2\\
&   &   & 4 & 1\\
&   &   &   & 5\\
\end{bmatrix},
F =
\begin{bmatrix}
- & \text{G} & \text{G} & \text{G} & \text{F}\\
& - & \text{N} & \text{N} & \text{F}\\
& & - & \text{F} & \text{F}\\
& & & - & \text{F}\\
& & & & - \\
\end{bmatrix},
P =
\begin{bmatrix}
- & 2.34 & 1.74 & 2.46 & 5.85\\
& - & 0.48 & 0.44 & 2.89\\
& & - & 1.47 & 1.29\\
& & & - & 0.24\\
& & & & - \\
\end{bmatrix}.
\]

The vine array, bivariate copula family array, and bivariate copula parameter array for model (E) in Section~\ref{subsec: vine_simulation} are as follows:
\[
A = 
\begin{bmatrix}
5 & 5 & 3 & 3 & 3 & 3 & 2 & 5\\
& 3 & 5 & 5 & 5 & 5 & 3 & 3\\
&   & 7 & 7 & 7 & 2 & 5 & 2\\
&   &   & 2 & 2 & 7 & 7 & 1\\
&   &   &   & 6 & 6 & 6 & 7\\
&   &   &   &   & 1 & 1 & 6\\
&   &   &   &   &   & 4 & 4\\
&   &   &   &   &   &   & 8\\
\end{bmatrix},
F =
\begin{bmatrix}
- & \text{G} & \text{G} & \text{G} & \text{G} & \text{G} & \text{G} & \text{F}\\
& - & \text{G} & \text{G} & \text{G} & \text{G} & \text{G} & \text{F}\\
&   & - & \text{N} & \text{N} & \text{N} & \text{N} & \text{F}\\
&   &   & - & \text{N} & \text{N} & \text{N} & \text{F}\\
&   &   &   & - & \text{F} & \text{F} & \text{F} \\
&   &   &   &   & - & \text{F} & \text{F} \\
&   &   &   &   &   & - & \text{F}\\
&   &   &   &   &   &  & - \\
\end{bmatrix},
\]
\[
P =
\begin{bmatrix}
- & 1.67 & 1.84 & 1.76 & 1.76 & 1.52 & 1.62 & 4.82 \\
& - & 1.45 & 1.26 & 1.26 & 1.42 & 1.47 & 2.19 \\
&   & - & 0.25 & 0.14 & 0.30 & 0.26 & 1.08 \\
&   &   & - & 0.13 & 0.18 & 0.18 & 0.92 \\
&   &   &   & - & 0.29 & 0.41 & 0.47 \\
&   &   &   &   & - & 0.13 & 0.23 \\
&   &   &   &   &   & - & 0.45 \\
&   &   &   &   &   &   & - \\
\end{bmatrix}.
\]

\section{Fitted Vine Copula Model for the PBC Dataset}
The fitted vine array and bivariate copula family array selected based on our proposed method in Section~\ref{subsec: selection} for the training data of the PBC dataset are as follows:
\[
A = 
\begin{bmatrix}
1 &1 &2 &7 &2 &7 &2 &2\\
0 &2 &1 &2 &7 &4 &7 &3\\
0 &0 &7 &1 &4 &2 &3 &7\\
0 &0 &0 &4 &1 &1 &4 &6\\
0 &0 &0 &0 &3 &3 &1 &4\\
0 &0 &0 &0 &0 &5 &5 &1\\
0 &0 &0 &0 &0 &0 &6 &5\\
0 &0 &0 &0 &0 &0 &0 &8\\
\end{bmatrix}, \quad
F =
\begin{bmatrix}
- & \text{G.s} & \text{N} & \text{G.s} & \text{G.s} & \text{G.s} & \text{BB1} & \text{G.s}\\
& - & \text{G.s} & \text{G.s} & \text{F} & \text{G.s} & \text{G.s} & \text{F}\\
&   & - & \text{N} & \text{t} & \text{N} & \text{G.s} & \text{F}\\
&   &   & - & \text{N} & \text{F} & \text{N} & \text{N}\\
&   &   &   & - & \text{N} & \text{N} & \text{N} \\
&   &   &   &   & - & \text{G.s} & \text{G} \\
&   &   &   &   &   & - & \text{N}\\
&   &   &   &   &   &  & - \\
\end{bmatrix}.
\]
The variables are (1) age, (2) edema, (3) bili, (4) albumin, (5) platelet, (6) protime, (7) stage, and (8) the response variable (survival time). N, t, F, G, G.s, and BB1 in the copula family matrix stand for bivariate Gaussian (normal), t, Frank, Gumbel, survival Gumbel, and BB1 copulas, respectively. Due to censoring, the upper tail behaviors of the response variable can only be partially observed. As a result, copulas with light upper tail are mostly chosen as the bivariate copula models associated with the survival time. 

\section{Examples of Comparing Different Prediction Methods Based on the PBC Dataset}
We randomly choose five samples from the test set of the PBC data and list the true observed or censored responses as well as the point and interval prediction results from the AFT and vine copula models for these samples in Table~\ref{tab: pbc_sample}. It can be seen that the vine copula model in general outputs smaller median predictions compared with the AFT model, which are often closer to the true survival times and lead to improved metrics.

\begin{table}[!ht]
	\small
	\centering
	\begin{tabular}{*{6}{c}}
		\toprule
		Survival time & Censored & Copula prediction & Copula 50\% PI & AFT prediction & AFT 50\% PI \\\midrule
		1702 & Yes & 5140 & [3845, 6613] & 5911 & [4007, 7974] \\
		4256 & Yes & 5885 & [4382, 7342] & 7134 & [4862, 9596] \\
		2400 & No & 2527 & [1612, 3564] & 2745 & [1795, 3774] \\
		51 & No & 129 & [50, 306] & 179 & [1, 371] \\
		1349 & Yes & 3784 & [2539, 5397] & 3863 & [2576, 5257] \\
		\bottomrule
	\end{tabular}
	\caption{The true observed or censored survival times, censoring status, median predictions and the corresponding 50\% prediction intervals generated by the AFT and vine copula models for five randomly selected samples from the test set.}
	\label{tab: pbc_sample}
\end{table}

\end{document}